\documentclass[useAMS,usenatbib]{mn2e}
\usepackage{graphicx}
\usepackage{rotating}

\title[CCD Photometry of NGC 5053]{CCD time-series photometry of
 the globular cluster NGC~5053: RR Lyrae, Blue Stragglers and SX Phoenicis stars revisited\thanks{Based on
  observations collected at the Indian
   Astrophysical Observatory, Hanle, India.}}
\author[A. Arellano Ferro et al.]{A. Arellano Ferro$^{1}$, Sunetra Giridhar$^{2}$, D.M. Bramich$^{3}$\\
$^{1}$Instituto de Astronom\1a, Universidad Nacional Aut\'onoma de M\'exico, M\'exico: armando@astroscu.unam.mx\\
$^{2}$Indian Institute of Astrophysics, Koramangala 560034, Bangalore, India: giridhar@iiap.res.in\\
$^{3}$European Southern Observatory, Karl-Schwarzschild-Stra\ss{}e 2, 85748 Garching bei M\"unchen, Germany: dan.bramich@hotmail.co.uk\\}

\begin{document}
 
\date{Accepted . Received ; in original form }

\pagerange{\pageref{firstpage}--\pageref{lastpage}} \pubyear{2002}

\maketitle 

\label{firstpage}

\begin{abstract}

We report the results of CCD $V$, $r$ and $I$ time-series photometry of the globular cluster NGC 5053. New times of maximum light are given
for the  eight known RR Lyrae stars
in the field of our images and their periods are revised.
Their $V$ light curves were Fourier decomposed to estimate their physical parameters. A discussion on the accuracy of the Fourier-based iron abundances, temperatures, masses and radii is given. New periods are found for the 5 known SX Phe stars and a critical discussion of their secular period changes is offered. The mean iron abundance for the RR Lyrae stars is found to be [Fe/H] $\sim -1.97 \pm 0.16$ and lower values are not supported by the present analysis. The absolute magnitude calibrations 
of the RR Lyrae stars yield an average true distance modulus of $16.12 \pm 0.04$ or a distance of $16.7 \pm 0.3$ kpc. Comparison
of the observational CMD with theoretical isochrones indicates 
an age of  $12.5 \pm 2.0$ Gyrs for the cluster. A careful identification of all reported Blue Stragglers (BS) and their $V,I$ magnitudes leads to the conclusion that BS12, BS22, BS23 and BS24 are not BS.  On the other hand, three new BS are reported.  Variability was found in seven BS, very likely of the SX Phe type in five of them, and in one red giant star. The new SX Phe stars follow established $PL$ relationships and indicate a distance in
agreement with the distance from the RR Lyrae stars.

\end{abstract}
      

\section{Introduction}

Recent numerical methods in the analysis of stellar CCD images have proven to be very powerful
in isolating faint stars and performing accurate photometry even in very crowded fields, such
as in the central regions of globular clusters (Bramich 2008, Bramich et al. 2005; 
Alard 2000; Alard \& Lupton 1998).
In a series of papers we have taken advantage of these mathematical techniques to study the known variable stars and to search for new ones in several globular clusters of assorted metallicities
(Arellano Ferro et al. 2008a; 2008b; 2004, L\'azaro et al. 2006). 
In particular we used the Fourier decomposition technique to estimate fundamental physical parameters of the RR Lyrae stars and we searched for eclipsing binaries and variability among the blue stragglers. The calculation of the iron content [Fe/H] and absolute magnitude $M_V$ lead to a linear $M_V$-[Fe/H] relationship (Arellano Ferro et al. 2008b) that is in agreement with 
independent estimations from different empirical techniques (e.g. Cacciari \& Clementini 2003; 
Chaboyer 1999).

The globular cluster NGC 5053 (R.A.(J2000)$=13^h16^m27^s$.3, DEC(J2000)$=+17^{\circ} 41'52''$, $l=335.7$, $b=+78.9$) is located in the intermediate Galactic halo ($Z=$ 16.1 kpc, R$_G$=16.9 kpc) and it is one of the most metal deficient clusters; [Fe/H] = $-2.1$ (Rutledge et al. 1997). 
It is a rather disperse and well resolved system which allows high quality photometry even in the central regions. Such characteristics place this stellar system close to the
grey region between globular clusters and open clusters. Nevertheless
it was classified as a globular cluster mainly due to its high galactic latitude and the presence of faint stars and some variables (Cuffey 1943, Baade 1928).
The cluster has been the subject of detailed photometric studies and the most recent ones, already in the CCD era, conducted detailed studies of known RR Lyrae stars (Nemec, Mateo \& Schombert 1995) and the identification of 28 blue stragglers (Nemec \& Cohen 1989; Sarajedini \& Milone 1995) among which 5 are known to be SX Phe variables (Nemec et al. 1995). 

In the present paper we perform standard $V$, $I$ and instrumental $r$ CCD photometry for nearly 6500 stars in the field of NGC 5053 and perform the Fourier 
light curve decomposition of the known RR Lyrae stars. We also search for variability among Blue Straggler (BS) stars and revisit the known SX Phe stars previously studied by Nemec et al. (1995), in an attempt to refine their periodicities in order to discuss their period changes and to use them as independent indicators of the distance to the cluster. Finally 
we report five new SX Phe stars, three new BS stars, and one new red giant variable (RGV). 

The paper is organised as follows:
in $\S$ 2 we describe the observations, data reductions and transformations to the standard photometric system. In $\S$ 3 a period analysis of the RR Lyrae stars is performed. In $\S$ 4 the $V$ light curves of the RR Lyrae stars are Fourier decomposed and their
physical parameters are estimated. In $\S$ 5
the periods and secular period changes of the SX Phe stars are discussed as well as the
BS nature of previously reported BS stars. In $\S$ 6 a search for new variables is carried out and the new variables are described. In $\S$ 7
the distance and age of NGC~5053 are discussed and in $\S$ 8 we summarise our conclusions.

\section{Observations and Reductions}
\label{sec:Observations}

The observations employed in the present work were performed using the Johnson $V$, $R$ and $I$ filters and
obtained with the 2.0m telescope of the Indian Astronomical Observatory (IAO), Hanle, India, located at 4500m above sea level, during several runs between April 2006 and January 2009. The cluster was observed during a total of 10 nights and 151 images were gathered in the Johnson $V$, 139 in $R$ and 13 in $I$. The average seeing was $\sim$1 arcsec.
The detector was a 
Thompson CCD of 2048 $\times$ 2048 pixels with a pixel
scale of 0.17 arcsec/pix and a field of view 
of approximately $11. \times 11.$ arcmin.

\begin{figure} 
\includegraphics[width=8.cm,height=8.cm]{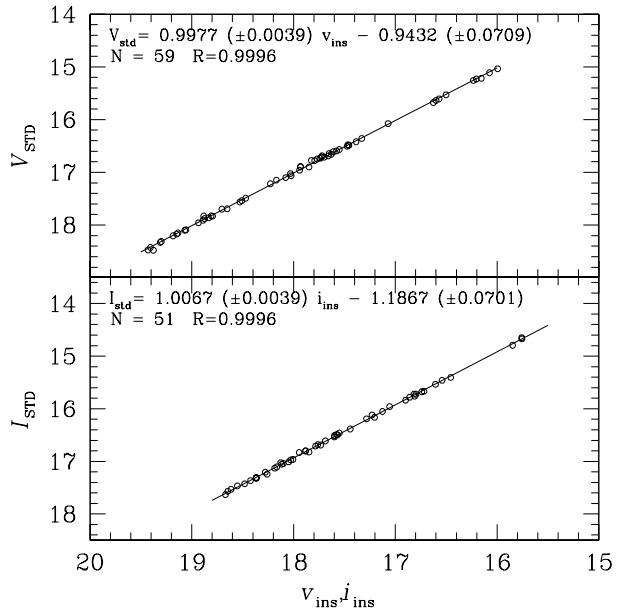}
\caption{Photometry transformation relations calculated using the standard stars in $V$ and $I$ from
the collection of Stetson (2000).}
    \label{2CAJAS}
\end{figure}

Difference image analysis (DIA) is a powerful technique allowing
accurate PSF photometry of CCD images, even in very crowded fields
(Alard \& Lupton 1998; Alard 2000; Bramich et al. 2005).
In the present study, we used a pre-release version of the 
{\tt DANDIA}{\footnote{
{\tt DANDIA} is built from the DanIDL library of IDL routines available at http://www.danidl.co.uk}}
software for the DIA
(Bramich et al., in preparation), which employs a new algorithm for
determining the convolution kernel matching a pair of images of the same
field (Bramich 2008). This algorithm was applied to a set of $V$, $R$ and $I$
images of NGC 5053 in order
to obtain accurate time-series photometry and to search for new variable stars down to $V \sim 19.5$ mag.

For a brief description of the DIA
procedure the reader is referred to $\S$ 2 of the paper by Arellano Ferro et al. (2008b)

\begin{table}
\footnotesize{
\begin{center}
\caption[{\small }] {\small Periods of the known RR Lyrae stars in NGC~5053. Most periods have been revised in this work (column 2) and new times of
maximum light were calculated (column 3). The predicted periods using the periods and period
change rates of Nemec (2004) are listed in column 4.}
\label{newephem}
\hspace{0.01cm}
 \begin{tabular}{llcc}
\hline
 ID & Period & Times of Maximum & Predicted Period  \\
& (days) & (+240~0000)& (days)\\
\hline
V2 & 0.378960& 53832.325 & 0.378956 \\
   &         & 53862.270 & \\
V3 & 0.592950& 53862.158 & 0.592946\\
V4 & 0.667073$^1$& 54202.326  &0.667079\\
V5 & 0.714860$^1$& 53833.300&0.714866\\
V6 & 0.292185& 53832.390 & 0.292178\\
   &         & 53833.265 &\\
   &         & 53861.323&\\
   &         & 54201.400&\\
V7 & 0.351900& 53828.721& 0.351927\\
V8 & 0.362864& 53833.259&0.362857\\
V10& 0.77585$^1$ & 53833.259 &0.775850\\
\hline
\end{tabular}
\end{center}
1. Nemec (2004)\\
}
\end{table}

\begin{figure*} 
\includegraphics[width=16.cm,height=10.cm]{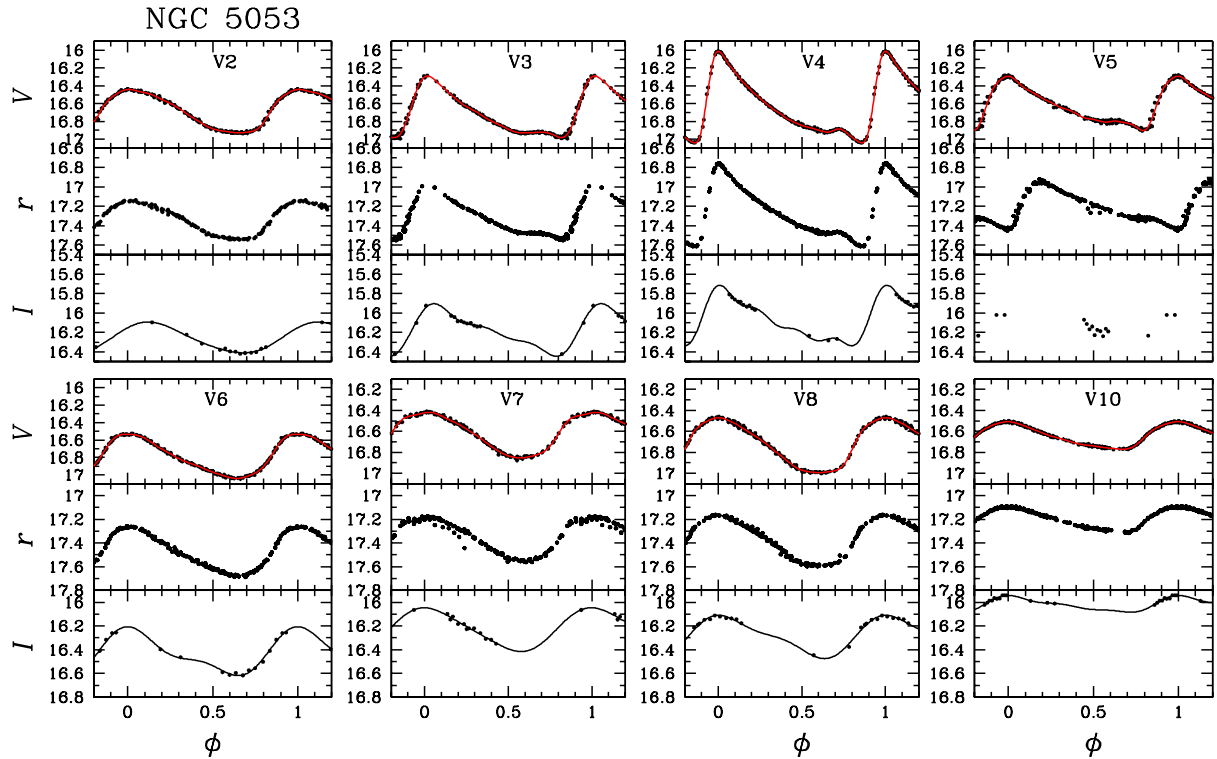}
\caption{$V, r, I$ light curves of the RR Lyrae stars in NGC~5053 phased with the new ephemerides from column 2 in Table \ref{newephem}.}
    \label{RRLC}
\end{figure*}

The instrumental $v$ and $i$ magnitudes were converted to the  Johnson $V$
and $I$ standard system using the collection of standards of Stetson (2000) (http://www3.cadc-ccda.hia-iha.nrc-cnrc.gc.ca/community/STETSON/standards/).
For NGC 5053 we identified 59 $V$ and 51 $I$ standard stars in the field of the cluster and the transformations are shown in Fig. \ref {2CAJAS}. The transformations were found to be linear and no colour term was found to be significant. The linear correlation coefficients were of the order of 0.999.
The instrumental $r$ magnitudes were retained in the instrumental system since no $R$ standards were found in the literature.

\begin{table*}
\footnotesize{
\begin{center}
\caption[Nuevas variables] {\small $V, r, I$ magnitudes of the RR Lyrae, variable BS and new variable stars in NGC~5053 (extract).}
\label{sample1}
\hspace{0.01cm}
\begin{tabular}{ccccccc}
\hline
Star & HJD$_V$ & $V$& HJD$_r$ & $r$& HJD$_I$ & $I$\\

\hline
V2&2453832.2276& 16.898 & 2453832.2374& 17.487&2454839.4620& 16.399\\ 
V2&2453832.2334& 16.876 & 2453832.2411& 17.484&2454839.4500& 16.395\\
V2&2453832.2484& 16.800 & 2453832.2438& 17.468&2454839.4272& 16.380\\ 
V2&2453832.2525& 16.772 & 2453832.2562& 17.390&2454839.4389& 16.365\\ 
V2&2453832.2626& 16.689 & 2453832.2589& 17.374&2454839.4724& 16.419\\	
...&...&...&...&...&...&...\\
\hline
\end{tabular}
\end{center}
}
\end{table*}

\section{The RR Lyrae stars}
\label{sec:RRLYR}

\subsection{New ephemerides and times of maximum light of the RR Lyrae stars}

Ten RR Lyrae stars are known in NGC 5053 (Clement et al. 2001) but only
eight are in the field of our collection of images. Our light curves were originally phased with the ephemerides of Nemec (2004). However, except for V5 and V10, we noticed that this produced poor quality folded light curves. To calculate 
the periods at the epoch of our observations we applied the phase dispersion minimisation technique (PDM)
(Burke et al. 1970; Dworetsky 1983). The periods that we found
are listed in column 2 of Table \ref{newephem}. In most cases the new periods differ substantially from thoes from Nemec (2004) (e.g. V3, V7 and V8) while for V5 and V10 the periods coincide within $3\times 10^{-6}$ days;  in these cases
the periods from Mannino (1963) and Nemec (2004) were adopted respectively. 
We have also estimated a number of epochs of maximum $V$ light and these are listed in column 3 of Table \ref{newephem}. 

In Fig.~\ref{RRLC} the light curves in standard $VI$ and instrumental $r$ magnitudes are displayed. The vertical scale is the same for all stars and filters so that amplitude and brightness differences can be seen at a glance.

All our $V$, $r$ and $I$ photometry for the RR Lyrae stars is available in Table \ref{sample1}. Only a small portion of this table is shown in the printed version but a full version is available in electronic form.

\subsection{Updated periods and period change rates for the RR Lyrae stars}

A thorough analysis of the secular variations in the periods of the RR Lyrae stars in NGC~5053 was carried 
out by Nemec (2004). Since only a few years have elapsed
between the most recent data of Nemec in 2002 and the earliest data from this paper in 2006,
there is very little that we could add to Nemec's O-C diagrams and/or the
 derived period change rates. We adopted his period change rates (column 11 of his 
Table 7) in order to predict the period at the average epoch of our data. The predicted periods are listed in column 4 of Table \ref{newephem}. These predicted periods do not phase our light curves correctly in all cases, and in some cases there are substantial differences with the best periods found in this work (column 2), which were
used to phase the light curves shown in  Fig.~\ref{RRLC}.  Therefore, it seems likely that the secular period changes quoted by Nemec (2004) are overwhelmed by observational inaccuracies.

\section{Fourier decomposition approach to the physical parameters of RR Lyrae stars}

The red continuous curve in each of the $V$ light curve panels of Fig.~\ref{RRLC} is the Fourier fit to the data and it is mathematically represented by an equation of the  form:

\begin{equation}
m(t) = A_o ~+~ \sum_{k=1}^{N}{A_k ~cos~( {2\pi \over P}~k~(t-E) ~+~ \phi_k ) },
\label{eqfou}
\end{equation}

\noindent
where $m(t)$ are magnitudes at time $t$, $P$ the period and $E$ the epoch. A linear
minimisation routine is used to fit the data with the Fourier series model, deriving
the best fit values of the amplitudes $A_k$ and phases $\phi_k$ of the sinusoidal harmonics. 
From the amplitudes and phases of the harmonics in eq. \ref {eqfou}, the Fourier parameters are 
defined as: $\phi_{ij} = j\phi_{i} - i\phi_{j}$, and $R_{ij} = A_{i}/A_{j}$.
The number of harmonics required for the proper fit of a given light curve depends on the quality of the data; basically on the scatter and the coverage of the light curve. We have fitted as many harmonics as possible as long as their amplitudes are significant. 
The magnitude-weighted mean magnitudes $A_0$ (or $(V)$), the amplitudes and phases of the first four harmonics, which are relevant to the physical parameter calibrations described below, and the total number of harmonics employed (NH) are listed in Table \ref{parametrosc}.

\begin{table*}
\footnotesize{
\begin{center}
\caption[{\small Par\'ametros de los ajustes Fourier calculados para las curvas
de luz en el filtro V de las estrellas tipo {\bf RRc} del c\'umulo globular NGC
5466. Donde n es el n\'umero de arm\'onicos usados para el ajuste.}] {\small Fourier fit parameters for the V light curves.\label{foufit}}
\label{parametrosc}
\hspace{0.01cm}
  \begin{tabular}{lccccccccccc}
 \hline
\hline
&&&&&RRc stars&&&&&&\\
\hline
ID & HJD & P & $A_0$  & $A_1$ & $A_2$ & $A_3$ & $A_4$ & $\phi _{21}$ & $\phi _{31}$ & $\phi _{41}$ & NH \\
& 2400000.0+ &  (days)   &$\sigma_{A_0}$ &$\sigma_{A_1}$ &$\sigma_{A_2}$ &$\sigma_{A_3}$ &$\sigma_{A_4}$ & $\sigma_{\phi _{21}}$& $\sigma_{\phi _{31}}$& $\sigma_{\phi _{41}}$\\
\hline
V2     & 53832.325 & 0.378955   & 16.697 &0.246 &0.045 & 0.025& 0.010 &4.596 &2.827 &1.660 & 8 \\
       &            &            &  0.001 &0.001 &0.001 & 0.001& 0.001 &0.030 &0.052 &0.126 &  \\
V6     & 53832.390   & 0.292185   & 16.805 &0.238 &0.066 & 0.014& 0.009 &4.295 &2.408 &0.524 & 6 \\
       &            &            &  0.001 &0.001 &0.001 & 0.001& 0.001 &0.019 &0.078 &0.124 &  \\
V7     & 53828.721  & 0.351900 & 16.635 &0.220 &0.031 & 0.012& 0.010 &4.895 &2.956 &1.873 & 6 \\
       &            &            &  0.001 &0.001 &0.001 & 0.001& 0.001 &0.040 &0.103 &0.124 &  \\
V8     & 53833.259  & 0.362864   & 16.747 &0.265 &0.048 & 0.022& 0.013 &4.595 &2.700 &1.699 & 7 \\
       &            &            &  0.001 &0.001 &0.001 & 0.001& 0.001 &0.031 &0.066 &0.113 &  \\
\hline
&&&&&RRab stars&&&&&&\\
\hline
V3     & 53826.573  & 0.592944   & 16.721 &0.280 &0.112 & 0.074& 0.036 &3.962 &1.877 &0.049 & 6 \\
       &            &            &  0.001 &0.002 &0.002 & 0.002& 0.002 &0.023 &0.035 &0.061 &  \\
V4     & 53822.093  & 0.667075   & 16.658 &0.352 &0.174 & 0.122& 0.084 &3.943 &1.905 &6.247 & 10 \\
       &            &            &  0.005 &0.001 &0.001 & 0.001& 0.001 &0.009 &0.013 &0.018 &  \\
V5     & 53829.726  & 0.7148605$^1$ & 16.639 &0.230 &0.100 & 0.065& 0.032 &4.078 &2.153 &0.514 & 8 \\
       &            &            &  0.002 &0.002 &0.002 & 0.002& 0.003 &0.033 &0.051 &0.086 &  \\
V10    & 53833.259  & 0.77585$^2$    & 16.653 &0.121 &0.034 & 0.009& 0.005 &4.190 &2.697 &2.091 & 6 \\
       &            &            &  0.001 &0.001 &0.001 & 0.001& 0.001 &0.023 &0.073 &0.134 &  \\
\hline
\end{tabular}
\end{center}
}
1. Mannino (1963), 2. Nemec (2004)
\end{table*}

In order to estimate the magnitude-weighted mean $(I)$, the $I$-curves were also analysed using the 
Fourier decomposition by eq. \ref{eqfou} despite possessing a sparse phase coverage.
The order of harmonics in this case was kept as low as possible until a smooth curve of  similar appearance to the $V$ light curve was obtained.
 
 The Fourier fit parameters in $V$ have been used to derive
 physical parameters of the RR Lyrae stars.
To calculate [Fe/H], and $M_V$ for each individual star we followed the procedure described in the
paper by Arellano Ferro et al. (2008b) where a detailed discussion on the Fourier approach,
its limitations and 
advantages is given. We do not repeat that discussion here and instead the interested reader is referred to that paper. In the following subsections we briefly describe the transformations employed to calculate the iron 
content and the luminosity for both RRc and RRab stars. Fourier based calculations of the effective temperatures, 
radii and masses are also addressed.

\subsection{The iron abundance [Fe/H]}

NGC~5053 has often been considered as the most metal deficient globular cluster in the Galactic Halo. Zinn (1985) lists the cluster with [Fe/H] = $-2.58\pm 0.27$ which corresponds to the value $-2.4\pm0.3$ in the Cohen scale determined by Bell \& Gustafsson (1983) via the comparison of synthetic spectra with the observed spectra of stars in the cluster. Despite the large uncertainty, this value for [Fe/H] is commonly found in the literature
for NGC~5053. Considerable effort has been invested in the calculation of the metallicity of the cluster after those first estimates, and numerous values are found in the literature: Suntzeff, Kraft \& Kinman (1988), from spectral analysis of the Ca H\&K found $-2.18 \pm 0.06$ and from the Mg 5130-5200\AA~ band: $-2.44 \pm 0.21$; Armandroff et al. (1992), from the analysis of the Ca II triplet in the 8500-8660\AA~ region found $-2.41 \pm 0.12$. A detailed discussion on the iron abundance of NGC~5053 has been 
given by Geisler et al. (1995) from which it can be highlighted that the more recent determinations find considerably higher values of [Fe/H]. Geisler et al. (1995), from an analysis of medium resolution spectra of the Ca II triplet, found [Fe/H]= $-2.10 \pm 0.06$, and by reconsidering $BVI$ data from Sarajedini \& Milone (1995)
and the Ca II triplet equivalent widths of Armandroff et al. (1992), find the values of $-2.21$ and $-2.22$ respectively. Finally, Geisler et al. conclude that a likely value of [Fe/H] in NGC~5053 is $-2.15$.
Geisler et al. (1995) also argue that the lower metallicity limit for globular clusters in the Galaxy is $-2.25\pm0.10$. A more recent analysis of the Ca II triplet by Rutledge et al. (1997) derives a value of
$\sim -2.10 \pm 0.07$ for the iron content of this cluster.

For the RRab stars, the calibration of Jurcsik \& Kov\'acs (1996) was employed:

\begin{equation}
	{\rm [Fe/H]}_{J} = -5.038 ~-~ 5.394~P ~+~ 1.345~\phi^{(s)}_{31},
\label{FEAB}
\end{equation}

\noindent
where $\phi^{(s)}_{31}$ is the phase in a sine series. The phase in a sine series $\phi^{(s)}_{31}$ is related to the phase in a cosine series $\phi^{(c)}_{31}$ via:
  ~~ $\phi^{(s)}_{jk} = \phi^{(c)}_{jk} - (j - k){\pi \over 2}$. The standard deviation of this calibration is 0.14 dex (Jurcsik 1998). 

The metallicity scale of the above equation was transformed into the ZW scale using
the relation  ${\rm[Fe/H]}_{J} = 1.431 {\rm[Fe/H]}_{ZW} + 0.88$ (Jurcsik 1995). 

Eq. \ref{FEAB} is applicable to RRab stars with a  {\it deviation parameter} $D_m$, defined by
Jurcsik \& Kov\'acs (1996) and Kov\'acs \& Kanbur (1998), not exceeding a proper limit.
These authors suggest $D_m \leq 3.0$. $D_m$ values for individual RRab stars are given in column 2 of Table ~\ref{fisicosAB}. All stars fulfill the condition except 
V10 which is not included in the [Fe/H] average.
The iron abundances [Fe/H]$_{ZW}$ for the RRab stars
are reported in Table \ref{fisicosAB}

\begin{table*}
\footnotesize{
\begin{center}
\caption[Parametros estelares de las RR Lyrae del tipo ab] {\small Physical parameters for the RR{\lowercase {ab}} stars.} 
\label{fisicosAB}
\hspace{0.01cm}
 \begin{tabular}{lccccccc}
\hline 
Star&$D_m$&[Fe/H]$_{J}$ &[Fe/H]$_{ZW}$ & $M_V(K)$ & log$(L/L_{\odot})$&  $\mu _0$  & $D$ (kpc)\\
\hline
 V3  &  1.5    &$-1.49$ &$-1.65$  & 0.57 & 1.711 & 16.091& 16.53 \\
 V4  &  1.5    &$-1.85$ &$-1.90$  & 0.43 & 1.727 & 16.168& 17.12 \\
 V5  &  0.8    &$-1.72$ &$-1.82$  & 0.47 & 1.712 & 16.111& 16.68 \\
 V10 & 11.7 &$-1.31^1$ &$-1.53^1$  & 0.48 & 1.706 & 16.111& 16.68 \\
\hline
average & &$-1.64$&$-1.76^2$ & 0.49  &1.714 & 16.120 &16.75\\
$\sigma$& &$\pm$0.18 &$\pm$0.13 & $\pm$0.06 & $\pm$0.009 &$\pm$0.033 & $\pm$0.26 \\
\hline
\end{tabular}
\end{center}
1: value not included in the averages.
2: this value is to be decreased by $\sim$~0.2-0.3 dex to bring the Fourier results in
agreement with spectroscopic values. See text for discussion.
}
\end{table*}

\begin{table*}
\footnotesize{
\begin{center}
\caption[Parametros estelares de las RR Lyrae del tipo  c] {\small  Physical parameters for the 
RR{\lowercase {c}} stars.}
\label{fisicosC}
\hspace{0.01cm}
 \begin{tabular}{lccccc}
\hline 
Star& [Fe/H]$_{ZW}$ & $M_V(K)$ & log$(L/L_{\odot})$ &$\mu _0$ &$D$ (kpc)\\
\hline
V2  & -2.11 & 0.52 & 1.700 & 16.120 & 16.75 \\
V6  & -1.71 & 0.62 & 1.664 & 16.127 & 16.80 \\
V7  & -1.98 & 0.53 & 1.698 & 16.045 & 16.18 \\
V8  & -2.09 & 0.52 & 1.702 & 16.168 & 17.12 \\
\hline
average  & -1.97 & 0.55 &1.691 &16.115 & 16.71 \\
$\sigma$ & $\pm$0.18 &$\pm$0.05 & $\pm$0.018 & $\pm$0.051& $\pm$0.39 \\
\hline
\end{tabular}
\end{center}
}
\end{table*}

Mean values [Fe/H]$_{ZW}$= $-1.97 \pm 0.18$ for the RRc stars and [Fe/H]$_{ZW}$= $-1.76 \pm 0.13$ for the RRab stars
were obtained. 
The problem of the Jurcsik \& Kov\'acs (1996) calibration of eq. \ref{FEAB} giving
values of [Fe/H] too large by $\sim$ 0.2-0.3 dex for very metal poor stars
has been discussed particularly for the RRab stars in NGC 5053 by Nemec (2004), whose result [Fe/H]=$-1.73$ is very similar to ours ($-1.76$). This issue has also been pointed out by
Jurcsik \& Kov\'acs (1996) themselves, Kov\'acs (2002) and for other clusters by  Cacciari et al. (2005) (M3), L\'azaro et al. (2006) (M2), Arellano Ferro et al. (2008a) (NGC~5466) and  Arellano Ferro et al. (2008b) (NGC~6366). It is also interesting to note that a good agreement between the Fourier  and spectroscopic [Fe/H] values for the LMC RR Lyrae stars was found by Gratton et al. (2004) and Di Fabrizio et al. (2005) after a systematic correction of $\sim~-0.2$ dex was applied to the Fourier values. Therefore, applying such correction to the mean value of [Fe/H] for the RRab one obtains [Fe/H]$_{ZW}$= $-1.96 \pm 0.13$.

For the RRc stars, we used the calibration of Morgan et al. (2007);

$$ {\rm [Fe/H]}_{ZW} = 52.466~P^2 ~-~ 30.075~P ~+~ 0.131~\phi^{(c)~2}_{31}  $$
\begin{equation}
~~~~~~~	~-~ 0.982 ~ \phi^{(c)}_{31} ~-~ 4.198~\phi^{(c)}_{31}~P ~+~ 2.424,
\end{equation}

\noindent
where $\phi^{(c)}_{31}$ is the phase in a series of cosines, and $P$ the period in days. This calibration 
provides iron abundances in the metallicity scale of Zinn \& West (1984) (ZW) with a standard deviation of 0.14 dex. The iron abundances [Fe/H]$_{ZW}$ for the RRc stars
are reported in Table ~\ref{fisicosC}.
We find a value of [Fe/H]$_{ZW}$= $-1.97 \pm 0.18$
in excellent agreement with the value derived from the analysis of the RRab stars after the correction discussed above.
The calibration of Morgan et al. (2007) has been applied to RRc stars in  M15 (-2.12)
(Arellano Ferro et al. 2006) and NGC 5466 (-1.92) (Arellano Ferro et al. 2008a) and has proven to yield values in very good agreement  with generally accepted and well established values from different techniques. 

In conclusion, the present results also support the fact that iron values obtained from the calibration of
eq. \ref{FEAB} for RRab stars need to be 
corrected by about $-0.2$ to $-0.3$ dex. Guided by the result for RRc stars, we corrected the iron value for the RRab stars by -0.2. The present result also indicates that while the cluster is 
certainly very  metal-poor, the extreme deficiency ascribed to it earlier is not supported.

\subsection{Visual absolute magnitudes and luminosities}

The luminosity, or absolute magnitude, for the RRc stars was calculated using the  empirical calibration of Kov\'acs (1998)

\begin{equation}
M_V(K) = 1.261 ~-~ 0.961~P ~-~ 0.044~\phi^{(s)}_{21} ~-~ 4.447~A_4,  
\label{MVRC}	
\end{equation}

\noindent
with a standard deviation of 0.042 mag. 

To bring the scale of eq. \ref{MVRC} into agreement with the mean magnitude for the RR Lyrae stars in the LMC, $V_0 = 19.064 \pm 0.064$ (Clementini et al. 2003), it has been necessary to decrease the zero point by 0.2$\pm$0.02 mag (Cacciari et al. 2005).
The reported $M_V$ values in Table \ref{fisicosC} have been calculated with a zero point of eq. \ref{MVRC} of 1.061, and are therefore in agreement with a distance modulus of the LMC of $18.5 \pm 0.1$ (Freedman et al. 2001; van den Marel et al. 2002; Clementini et al. 2003). If the $M_V$ values in Table \ref{fisicosC} are compared with the ones obtained by Nemec (2004) for the same stars after we subtract 0.2 mag from Nemec's results (labelled as K98 in his Table 14), it can be seen that the two sets of results agree to within $\sim$ 0.02 mag.

For the RRab stars we have used the calibration of Kov\'acs \& Walker (2001): 

\begin{equation}
M_V(K) = ~-1.876~log~P ~-1.158~A_1 ~+0.821~A_3 +K,
\label{MVAB}
\end{equation}

\noindent
which has a standard deviation of 0.04 mag.

The zero point of eq. \ref{MVAB}, $K=0.43$, has been calculated by Kinman (2002) using the star RR Lyrae as a calibrator adopting for RR Lyrae the absolute magnitude $M_V= 0.61 \pm 0.10$ mag, as derived by Benedict et al.  (2002) using the star parallax measured by the HST. If one adopts $<V>=19.064$ for the RR Lyraes in the LMC and a value of [Fe/H]=$-1.5$ (Clementini et al. 2003), then correcting to [Fe/H]=$-1.39$
for RR Lyrae (Benedict et al. 2002) one obtains $<V>=19.088$. Hence, the value $K=0.43$ corresponds to a distance modulus of 19.088 $-$ 0.61= 18.48 mag. To make our calculations
of the luminosities for the RRab stars consistent with the above values and the distance modulus of 18.5 for the LMC (Freedman et al. 2000), we adopted $K=0.41$ in the application of eq. \ref{MVAB}.
The $M_V(K)$ values for the RRab stars are given in Table ~\ref{fisicosAB}. The average $M_V(K)$
is $0.49 \pm 0.06$ mag. This result is, within the observational scatter, in good agreement with the mean absolute magnitude found for the RRc stars of $0.55 \pm 0.05$ mag.
  
The values of $M_V(K)$ in Tables~\ref{fisicosAB} and \ref{fisicosC} were transformed into $log~L/L_\odot$. The bolometric corrections for the average temperatures of RRc and RRab stars  were estimated from the calibration for metal poor stars of Montegriffo et al. (1998).
We adopted the value $M_{bol}^{\odot} = 4.75$.

\subsection{The effective temperature $T_{\rm eff}$}
\label{secTeff}

The effective temperature of RRc and RRab stars can also be calculated from the Fourier decomposition technique. However, some problems have been identified with the calibrations of RRc stars as we shall discuss below.

For the RRc stars, the calibration of Simon \& Clement (1993) can be used:

\begin{equation}
	log T_{\rm eff} = 3.7746 ~-~ 0.1452~log~P ~+~ 0.0056~\phi^{(c)}_{31}.
\label{TEFSC}
\end{equation}

For the RRab stars we used the calibration of Jurcsik (1998)

\begin{equation}
	log~T_{\rm eff}= 3.9291 - 0.1112~(V - K)_o - 0.0032~[Fe/H]
\label{TefVK}
\end{equation}

\noindent
with 

$$ (V - K)_o= 1.585 ~+~ 1.257~P ~-~ 0.273~A_1 ~-~ 0.234~\phi^{(s)}_{31}  $$
\begin{equation}
~~~~~~~ ~+~ 0.062~\phi^{(s)}_{41}.
\label{VKPAF}
\end{equation}

\noindent
Eq. \ref{TefVK} has a standard deviation of 0.0018 (Jurcsik 1998), but the uncertainty in $log~T_{\rm eff}$ is mostly set by the uncertainty in the colour 
from eq. \ref{VKPAF}. The error estimate on  $log~T_{\rm eff}$ is 0.003 (Jurcsik 1998).

It has been pointed out by Cacciari et al. (2005) that the temperatures computed from the above equations do not match the colour-temperature relations predicted by the temperature scales of Sekiguchi \& Fukugita (2000) or by the  evolutionary models of Castelli (1999). This was in fact corroborated by Arellano Ferro et al (2008a) who concluded that the Fourier based temperatures for
the RRc need to be reduced by as much as $\sim$ 300 K. In the present paper we explore the colour temperatures from our $V,I$ data and the VandenBerg, Bergbusch \& Dowler (2006) HB models, and we perform a comparison 
with the Fourier temperatures for both RRc and RRab stars.

\begin{table*}
\footnotesize{
\begin{center}
\caption[Parametros estelares de las RR Lyrae del tipo ab] {\small Effective temperatures and radii for the RRc and RRab stars.}
\label{TeffRadii}
\hspace{0.01cm}
 \begin{tabular}{lccccccc}
\hline 
Star & $(I)$ & $(V-I)_0$& $log~T_{\rm eff} $ & $log~T_{\rm eff}$ & $log~R/R_{\odot}$ & $log~R/R_{\odot} $ & $log~R/R_{\odot}$ \\
&  & & $(Fou)$ & $(VI)$ & $(Fou)$ & $(VI)$ & $(PRZ)$ \\
\hline
&  & & & RRc stars &  &  & \\
\hline
 V2  &16.261  &0.434 &  3.852 & 3.833 & 0.673 & 0.708 &0.712 \\
 V6  &16.433  &0.364 &  3.866 & 3.854 & 0.632 & 0.647 & 0.642 \\
 V7  &16.232  &0.379 &  3.857 & 3.850 & 0.655 & 0.673 & 0.692 \\
 V8  &16.289  &0.447 &  3.854 & 3.829 & 0.670 & 0.717 & 0.700 \\
\hline
&  & & & RRab stars &  &  & \\
\hline
 V3  &16.188  &0.580 &  3.803 & 3.793 & 0.769 & 0.789 & 0.747 \\
 V4  &16.087  &0.545 &  3.797 & 3.802 & 0.789 & 0.780 & 0.777 \\
 V5  &16.100: &0.545 &  3.790 & 3.802 & 0.795 & 0.772 & 0.794 \\
 V10 &16.023  &0.609 &  3.781 & 3.786 & 0.812 & 0.802 & 0.815 \\
\hline
\end{tabular}
\end{center}
}
\end{table*}

The value of $A_0$ in eq. \ref{eqfou} and in Table \ref{parametrosc} corresponds in fact to the magnitude-weighted $(V)$ mean magnitude. The values of $(I)$ are listed in Table \ref{TeffRadii}. To deredden the mean magnitudes we have adopted $E(B-V)=0.018$ (Nemec 2004) and have calculated the colour ratio $E(B-V)/E(V-I) = 1.58$ from the colour ratios given by Barnes, Evans  \& Moffett (1978).
The values of $(V-I)_0$ for the sample RR Lyraes are also given in Table \ref{TeffRadii}.

To convert the observed $(V-I)_0$ to effective temperatures we have followed Nemec (2004) in the use of the  HB models of VandenBerg, Bergbusch \& Dowler (2006) with the colour-$log~T_{\rm eff}$ relations as described by VandenBerg \& Clem (2003).
The model with [Fe/H]=$-2.012$ and [$\alpha$ /H]=0.3 was adopted with a polynomial 
fit of the form:

\begin{equation}
y= A_0 + A_1 x + A_2 x^2+ A_3 x^3+ A_4 x^4+ A_5 x^5+ A_6 x^6+ A_7 x^7
\label{polynom}
\end{equation}

\noindent
where  y= $log~T_{\rm eff}$, x=$(V-I)_0$ and the coefficients $A_0=  3.9867, 
A_1= -0.9506, A_2= +3.5541, A_3= -3.4537, A_4= -26.4992, A_5=  +90.9507, A_6= -109.6680$ and $A_7=  +46.7704$. Lower order polynomials  do not reproduce the theoretical colour-$log~T_{\rm eff}$ relations well enough, and the selection of a model with 
[Fe/H]=$-2.310$ does not produce significantly different results. The temperatures obtained from
eq. \ref{polynom} are called $log~T_{\rm eff} (VI)$.

The values of the Fourier based temperatures, $log~T_{\rm eff} (Fou)$ (i.e. those obtained from 
eqs. \ref{TEFSC},  \ref{TefVK} and  \ref{VKPAF} for the RRc and RRab stars respectively) and the $VI$ colour temperatures, $log~T_{\rm eff} (VI)$, are listed in columns 4 and 5 of Table \ref{TeffRadii} respectively. These results indicate that for the RRc stars the Fourier based  temperatures are on average $\sim 255 K$ hotter than the  $VI$ colour temperatures, whereas for the RRab stars the agreement is better with the Fourier based temperatures being on average $\sim 40 K$ cooler.

\begin{figure} 
\includegraphics[width=8.cm,height=7.cm]{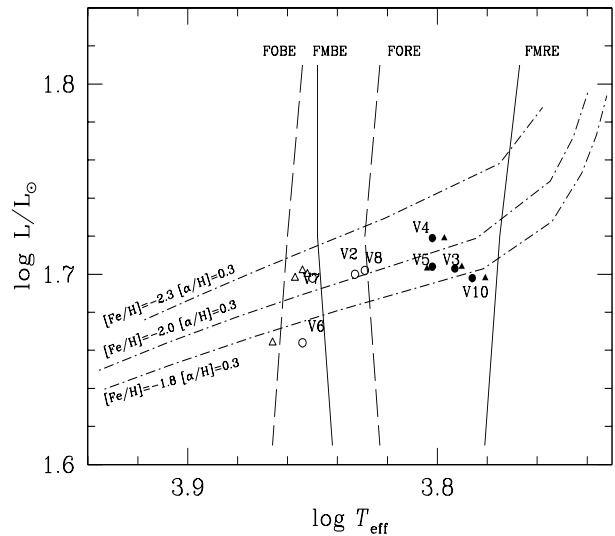}
\caption{RR Lyrae stars in the HRD for NGC~5053. Solid symbols represent RRab stars and open symbols RRc stars. The triangles were plotted using the Fourier based temperatures, while the circles correspond to colour temperatures calculated from the $(V-I)_0$ index. The vertical boundaries are the fundamental mode 
(continuous lines) and first overtone (dashed lines) instability strips from Bono et al. (1995) for 0.65 $M/M_\odot$.
 The models of the ZAHB (VandenBerg, Bergbusch \& Dowler 2006) are shown (dot-dashed lines) for [Fe/H]=$-1.836$, $-2.012$ and $-2.310$,
all for [$\alpha$/Fe]=+0.3. }
    \label{HRD}
\end{figure}

The distribution of the RR Lyrae stars in the instability strip is shown in Fig. \ref{HRD}.
Two positions have been plotted for each star; for $log~T_{\rm eff} (Fou)$ (triangles)
and $log~T_{\rm eff} (VI)$ (circles).
Three models of the Zero Age Horizontal Branch (ZAHB) from VandenBerg, Bergbusch \& Dowler (2006) for [Fe/H]=$-1.836$, 
[Fe/H]=$-2.012$ and [Fe/H]=$-2.310$ all for [$\alpha$/Fe]=+0.3 are plotted. The instability strip borders for the fundamental mode and first overtone are also shown (Bono et al. 1995).
The distribution of the RRc and the RRab stars resulting from the use of 
$log~T_{\rm eff} (Fou)$ values (triangles)
looks clumpy. The RRc stars are too blue and the large gap between the RRc and RRab stars is contrary to observational evidence in clusters with larger populations of RR Lyrae stars. For instance, in M3, the distribution of RR Lyrae stars across the instability strip is more even, and some fundamental mode and first overtone pulsator stars share  the inter-mode region (Cacciari et al. 2005).
The use of the colour temperatures $log~T_{\rm eff} (VI)$ (circles)
produces a larger spread in the RRc stars but not 
in the RRab stars, which indicates that the Fourier based temperatures for the RRab stars are consistent with 
the colour temperatures. The colour temperatures distribute the RRc stars more evenly accross the instability strip.

\subsection{The radius $R/R_\odot$}
\label{secRadii}

Given the stellar luminosity and the effective temperature, the stellar radius can be estimated  via the expression $log~R/R_\odot= [log~(L/L_\odot)-4 log~(T_{\rm eff}/T_{\rm eff}{}_\odot)]/2$.
With the Fourier decomposition based values of $log(L/L_\odot)$ (or $M_V$ for RRab) and $T_{\rm eff} (Fou)$, one can derive the stellar radii $log (R/R_{\odot}) (Fou)$. 
These radii depend fully on the semi-empirical relations and the hydrodynamical models used to calculate the
 luminosity and the temperature. 

An alternative calculation of the radii can be performed using the colour temperatures $T_{\rm eff} (VI)$. These radii we shall denote as $log (R/R_{\odot}) (VI)$.

A yet completely independent approach to the RR Lyrae radii determination is through the 
Period-Radius-Metallicity (PRZ) calibrations of Marconi {\it et al.} (2005) which are  
based on nonlinear convective models (e.g. Bono et al. 2003). Two calibrations are offered; for the first overtone pulsators or RRc stars: log $R/R_{\odot} = 0.774 + 0.580$ log $P - 0.035$ log Z,
and for the fundamental pulsators or RRab stars: log $R/R_{\odot} = 0.883 + 0.621$ log 
$P - 0.0302$ log Z.
We converted  the individual values of [Fe/H]$_{ZW}$ into Z making use of the equation: 
$log~\rm{Z} = \rm{[Fe/H]} - 1.70 + log (0.638~f + 0.362)$, where $f$ is the
 $\alpha$-enhancement factor with respect to iron (Salaris et al. 1993) which we adopt as
  $f=1$. These radii are listed in Table \ref{TeffRadii} as $log (R/R_{\odot})$ (PRZ) along with
the values of $log (R/R_{\odot}) (Fou)$ and $log (R/R_{\odot}) (VI)$. 

\begin{figure} 
\includegraphics[width=8.cm,height=8.cm]{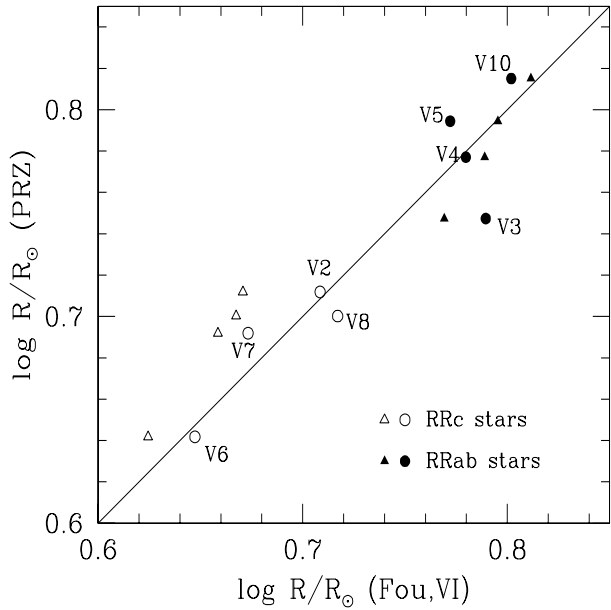}
\caption{Comparison of radii log $R/R_{\odot}$ (Fou) (triangles) calculated from the Fourier decomposition parameters log$(L/L_{\odot})$ and $T_{\rm eff}$,
with those obtained from the Period-Radius-Metallicity calibrations of Marconi {\it et al.} (2005), log $R/R_{\odot}$ (PRZ). The radii log $R/R_{\odot}$ (VI) calculated from the colour temperatures (circles) are in better agreement with log $R/R_{\odot}$ (PRZ)
particularly for the RRc stars.  Solid symbols represent RRab stars and open symbols RRc stars. See text in $\S$ \ref{secRadii} for discussion.}
    \label{RADII} 
\end{figure}

A comparison of all of the above estimates of the radii is summarised in Fig. \ref{RADII} where
open triangles indicate that the radii from the Fourier based temperatures, $log (R/R_{\odot}) (Fou)$ for the RRc stars are on
average 6\%  smaller than the PRZ radii $log (R/R_{\odot}) (PRZ)$. For the RRab 
stars, $log (R/R_{\odot}) (Fou)$  and $log (R/R_{\odot}) (PRZ)$ agree within their uncertainties (solid triangles).  On the other hand, the 
radii from the colour temperatures, $log (R/R_{\odot}) (VI)$, 
are in agreement with $log (R/R_{\odot}) (PRZ)$ for both the RRc and RRab stars
(circles). These results add a word of caution for the temperatures obtained from the calibrations of eq. \ref{TEFSC}, but confirm that the temperatures for the RRab stars from eqs.  \ref{TefVK} and  \ref{VKPAF} seem consistent with theoretical predictions as pointed out in $\S$ \ref{secTeff}.

\subsection{The mass $M/M_{\odot}$}
\label{secMass}

Fourier-based masses of the RRc stars can be estimated by the calibration of Simon \& Clement (1993):

\begin{equation}
log~M/M_\odot = 0.52~log~P ~-~ 0.11~\phi^{(c)}_{31} ~+~ 0.39,
\label{massRRc}
\end{equation}

\noindent
and for the RRab stars by the calibration of Jurcsik (1998);

$$ log~M/M_\odot =20.884 ~-~ 1.754~log~P ~+~ 1.477~log~(L/L_{\odot}) $$

\begin{equation}
~~~~~ ~-~ 6.272~log~T_{\rm eff} ~+~ 0.367~[Fe/H].
\label{massRRab}
\end{equation}

Cacciari et al. (2005) have argued that the masses obtained for RR Lyrae stars from the above relations and the Fourier decomposition of their light curves
are not reliable. In order to test these relations in the same way as for the effective temperature, one can compare these masses with those predicted by the fundamental equation of stellar pulsation of van Albada \& Baker (1971):

$$ log~M/M_\odot = 16.907 - 1.47~log~P_F + 1.24~log~(L/L_{\odot}) $$

\begin{equation}
- 5.12~log~T_{\rm eff}
\label{vanALBADA}
\end{equation}

\noindent
where $P_F$ is the fundamental period. In order to apply this equation to the first overtone pulsators RRc, their periods were transformed to the fundamental mode by adopting the ratio $P_{1H}/P_F = 0.748$, which is the average ratio for the double mode RR Lyrae stars in M15 (Cox et al. 1983). Note that the difference between eq. \ref{vanALBADA} and the calibration of Jurcsik (1998) in eq. \ref{massRRab} is that the later includes the iron dependence, and therefore a comparison of the two 
relations is of interest. However, after the discussion in  $\S$ \ref{secTeff} and $\S$ \ref{secRadii} of the temperature scale, the major concern is 
how the mass is affected by the chosen temperature scale.

\begin{table}
\footnotesize{
\begin{center}
\caption[Parametros estelares de las RR Lyrae del tipo ab] {\small Masses for the RRc and RRab stars.}
\label{MassGrav}
\hspace{0.01cm}
 \begin{tabular}{lccc}
\hline 
Star&  $M/M_{\odot}$ & $M/M_{\odot}$ &  $M/M_{\odot}$  \\
 & $(Fou)$ & $(VI)$& $(Pul)$   \\
\hline
 &   RRc stars  && \\
\hline
 V2  & 0.72 & -- & 0.67  \\
 V6  & 0.70 & -- & 0.69  \\
 V7  & 0.67 & -- & 0.60  \\
 V8  & 0.73 & -- & 0.75  \\
\hline
 &   RRab stars &&   \\
\hline
 V3  & 0.77 & 0.89 & 0.86 \\
 V4  & 0.70 & 0.65 & 0.68 \\
 V5  & 0.66 & 0.55 & 0.59 \\
 V10 & 0.66 & 0.61 & 0.62 \\
\hline
\end{tabular}
\end{center}
}
\end{table}

The values of the masses are reported in Table \ref{MassGrav}. For the RRc stars the Fourier mass $M/M_{\odot} (Fou)$ (eq. \ref{massRRc}) and the pulsational mass $M/M_{\odot} (Pul)$ (eq. \ref{vanALBADA}) agree within 10\%. Since $M/M_{\odot} (Pul)$ depends on the temperature, we calculated them using the colour temperatures $log~T_{\rm eff} (VI)$. However, if the 
Fourier temperatures $log~T_{\rm eff} (Fou)$ are used instead, the resulting Fourier masses would be 
more than 20\% too small. This also demonstrates that the Fourier temperatures for the RRc (eq. \ref{TEFSC}) are inaccurate.  

For the RRab stars the Fourier mass $M/M_{\odot} (Fou)$ (eq. \ref{massRRab} with Fourier temperature), the colour-temperature based masses $M/M_{\odot} (VI)$ (eq. \ref{massRRab} with colour temperature), and the pulsational mass $M/M_{\odot} (Pul)$ (eq. \ref{vanALBADA} with colour temperature) are all listed in Table \ref{MassGrav}. We notice that although the largest differences between the average of colour-temperature based masses 
($M/M_{\odot}$(VI) and  $M/M_{\odot}$(Pul)) minus the Fourier 
masses ($M/M_{\odot}$(Fou)) are +14\% for V3 and $-15$\% for V5, the average difference including the four RRab stars is only $-3$\%.

In conclusion, the Fourier temperatures for the RRc stars (from eq. \ref{TEFSC} of
Simon \& Clement (1993) seem to be $\sim 250K$ too hot relative to the 
colour-temperature calibration based on calculations from theoretical models of the HB. This is confirmed by the temperature-dependent Fourier radii and masses when compared with their pulsational estimates. This has also been
commented by Nemec (2004) for the RRc temperatures. However, contrary to 
the suggestion by Cacciari et al. (2005), and despite the larger scatter in the difference between the colour-temperature masses and the Fourier masses compared to the case for the radii,
we do not find evidence that the Fourier temperatures for RRab stars from equations  \ref{TefVK} and \ref{VKPAF} are significantly different from the
 colour temperatures. The radii and masses based on the Fourier temperatures are comparable to their pulsational estimates.  

\section{The Blue Stragglers and SX Phoenicis stars}
\label{secSXPHE}

\subsection{Periods of the SX Phe stars}
\label{PERSXPHE}

Four SX Phoenicis (SX Phe) in NGC 5053 were discovered by Nemec (1989) and a fifth
one was later found by Nemec et al. (1995).
Nemec et al. (1995) determined the main period and discussed the period changes. With our new collection of high quality light curves we believe that we are in a position to revisit these subjects.

\begin{figure*} 
\includegraphics[width=16.cm,height=16.cm]{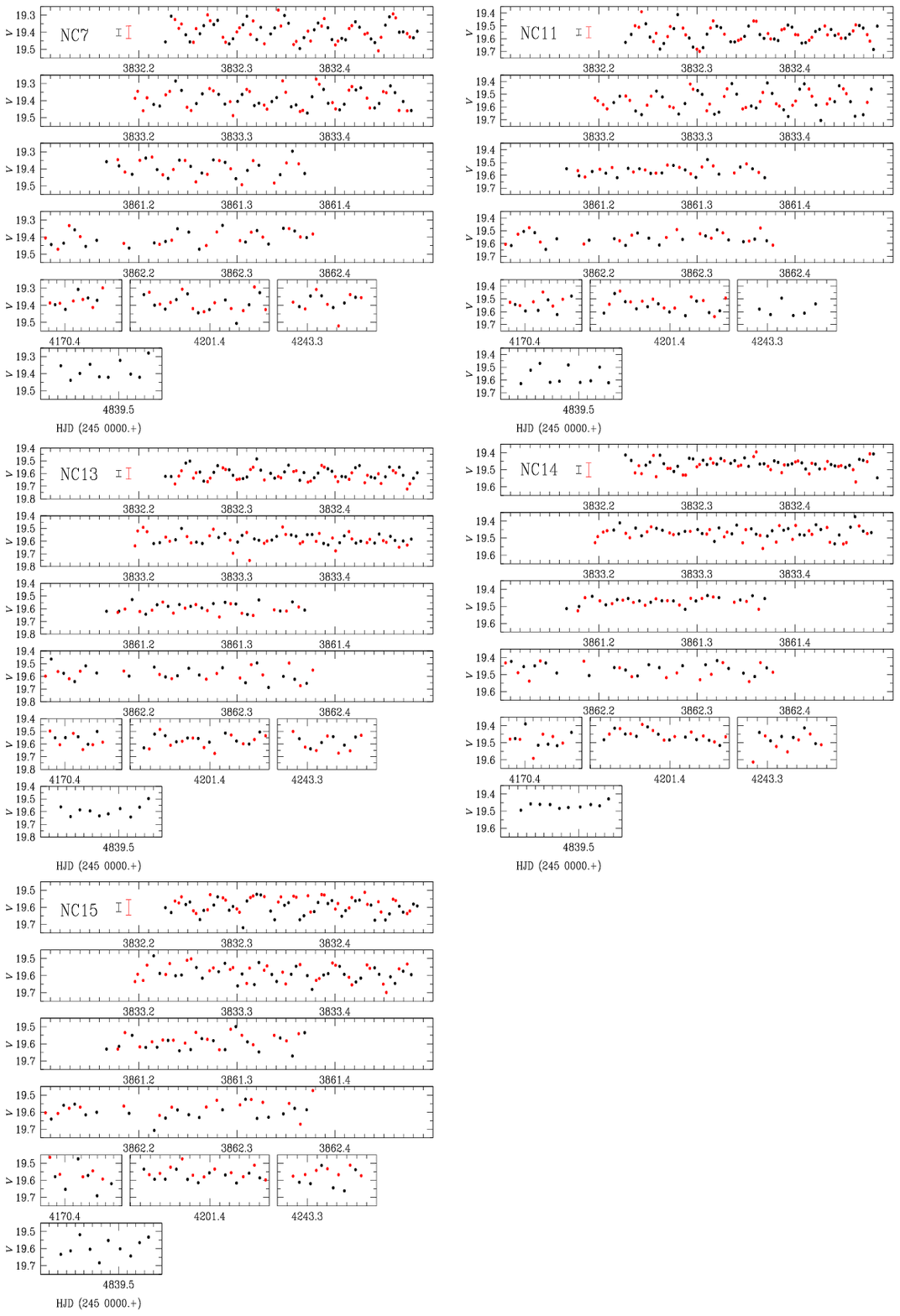}
\caption{$V$ (black circles) and $r$ (red circles) light curves of the known SX Phe stars in NGC~5053 as obtained in the present work. In order to highlight the variations, the $r$ light curves have 
been offset in magnitude such that the mean $r$ magnitude matches the mean $V$ magnitude for each star. 
Mean uncertainties for the $V$ and $r$ data points are plotted at the 
start of the light curve for clarity.}
    \label{SXstars}
\end{figure*}

The light curves of the SX Phe stars from our data set are displayed in Fig. \ref{SXstars}. We have used the Phase Dispersion Minimisation (PDM) approach (Burke et al. 1970; Dworetsky 1983) to first estimate the variability parameter SQ and the period. With the aim of confirming the PDM result and of searching for multiple frequencies we have also used the program
PERIOD04 (P4) (Lenz \& Breger 2005) on the present data for the five known SX Phe.
The final periods found by P4 are listed in column 3 of Table \ref{SXPHEper}. The uncertainties are $\sim 0.00001$ days. The periods found by Nemec et al. (1995) are also listed. The differences 
in period are real due to the intrinsic period variations of the stars and the 20-25 years elapsed between the two data sets. Due to the irregular nature of the variations, likely due to the presence of other undetected frequencies, the light curves in Fig. \ref{SXstars} are not folded with the periods listed in Table
\ref{SXPHEper}.

\begin{table}
\footnotesize{
\begin{center}
\caption[Nuevas variables] {\small PDM periods of the five known SX Phe stars.}
\label{SXPHEper}
\hspace{0.01cm}
\begin{tabular}{lcc}
\hline
Star & P(days) & P(days) \\
     & Nemec et al. (1995) & present work \\
\hline
NC7&   0.03683 & 0.03700 \\
NC11&  0.0350~~ & 0.03765 \\	
NC13&  0.03416 & 0.03396 \\	
NC14&  0.03925 & 0.03411 \\	
NC15&  0.0356~~ & 0.03574 \\	
\hline
\end{tabular}
\end{center}
}
\end{table}

\begin{table}
\footnotesize{
\begin{center}
\caption[] {\small Multiple frequencies in the SX Phe star NC13 in NGC~5053.}
\label{freqsSX}
\hspace{0.01cm}
 \begin{tabular}{lccc}
\hline 
 & NC13 &  &   \\
\hline
 & Frequency & Amplitude  &  Mode \\
    & (c/d)  &    (mag) & \\
\hline
$f_1$ & 29.44274 & 0.044& $F$ \\
 &$\pm 0.00005$ & & \\
$f_2$ & 28.85900 & 0.030& Non radial \\
 &$\pm 0.00007$ & & \\
$f_3$ & 37.31934 & 0.015& $1H$  \\
 &$\pm 0.00012$ & & \\
\hline
\end{tabular}
\end{center}
}
\end{table}

\begin{figure*} 
\includegraphics[width=15.cm,height=12.cm]{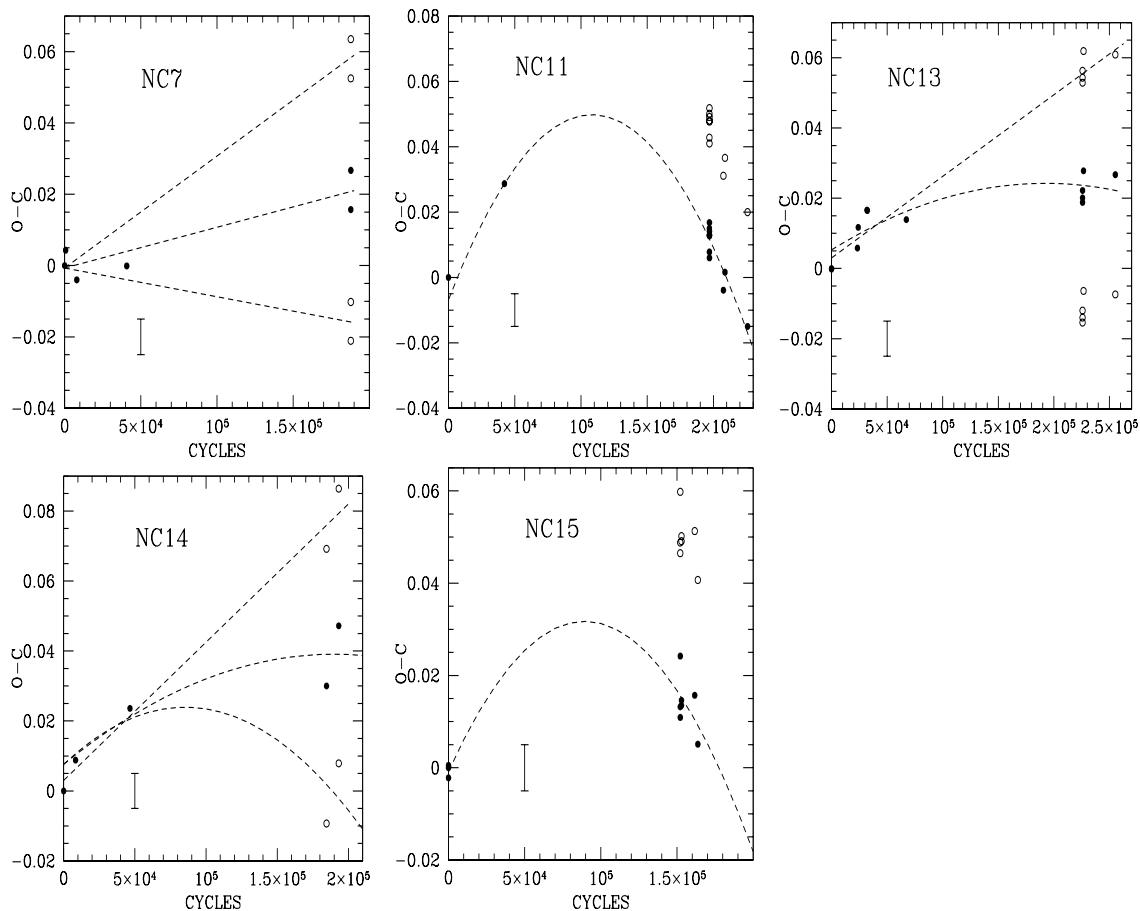}
\caption{O-C diagrams for the five known SX Phe stars in NGC~5053. Solid circles are the points used to calculate the most suitable parabolic fits, whereas open circles represent alternative values for certain key points by adding or subtracting one cycle. The error bar is the estimate of the uncertainty of each individual point.
See text for discussion.} 
    \label{OCDIAGS}
\end{figure*}

\begin{figure*} 
\includegraphics[width=17.cm,height=9.5cm]{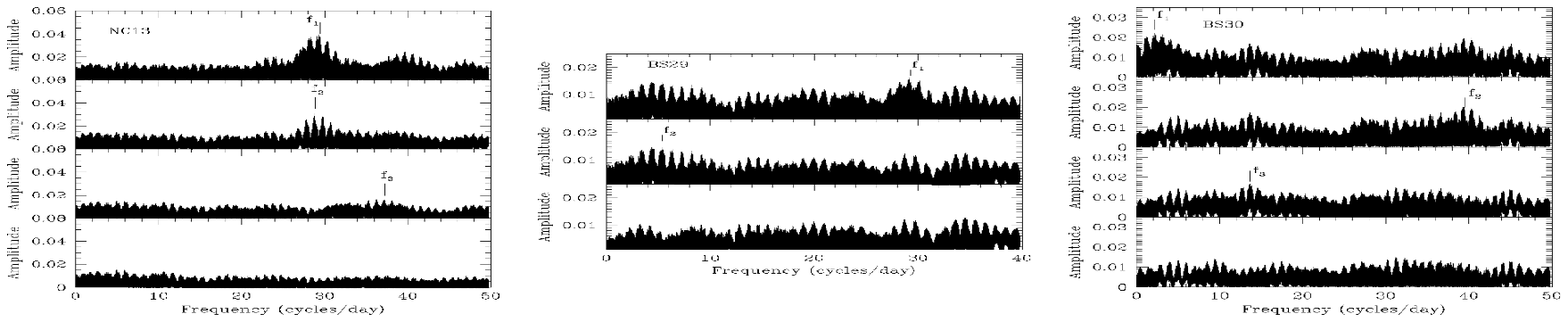}
\caption{Power spectra of NC13, BS29 and BS30}
    \label{POWERS}
\end{figure*}

Most likely due to the time distribution of our data set, we found only
one frequency in the known SX Phe except for NC13. 

For NC13 the period from the PDM method is 0.03421d. The frequencies from the PERIOD04 method are given in Table \ref{freqsSX} and correspond to the 
periods 0.03396d (in agreement with the PDM result), 0.03465d and 0.02680d. 
The power spectrum 
corresponding to these periods is shown in Fig. \ref{POWERS} (left). The main frequency and the third one are in the ratio $f_1/f_3 =0.789$ and therefore they 
are likely to be the fundamental mode and the first harmonic.

\subsection{Period changes in the SX Phe stars}

A first attempt to analyse the period changes in the SX Phe stars
in NGC~5053 was made by Nemec et al. (1995). Since nearly two decades have elapsed between the data employed by these authors and the present data, we have decided to revisit the subject. 

The times of maximum light employed by Nemec et al. (1995) are not explicitly given in their paper and we found that it was
inaccurate to try and recover them from their phase diagrams. Instead we opted
for inspecting their light curves and estimating the times of maximum.
The same procedure was performed on the present light curves in Fig.
\ref{SXstars}. The times of maximum are reported in Table \ref{timax}.

In order to build the O-C diagrams of Fig. \ref{OCDIAGS}, the periods of Nemec 
et al. (1995) and the first time of maximum were adopted as initial ephemerides.
Since the pulsational periods are very short and the time elapsed between the two data sets is quite large, the counting of cycles is sometimes uncertain. In Fig. \ref{OCDIAGS} the O-C values that we consider the most likely ones are plotted as
solid black circles but alternative values resulting from the addition or subtraction of one cycle are also plotted as open circles. The uncertainty 
of each individual data point was estimated by Nemec et al. (1995) to be about 0.1P. However, our estimations of several times of maximum from a single night show a larger dispersion indicated by the vertical
error bars in the figure. We believe this is a more conservative and realistic
estimate of the individual uncertainties. Brief comments on the period changes of each star are given below  

\begin{table*}
\footnotesize{
\begin{center}
\caption{\small Times of maximum light in the SX Phe stars.}
\label{timax}
\hspace{0.01cm}
\begin{tabular}{ccccc}
\hline
 NC7 & NC11 & NC13 & NC14 & NC15 \\
HJD&HJD&HJD&HJD&HJD\\
&&(2400000. +) &&\\
\hline
46914.995 & 46938.813 &46118.153 &46587.803 &48416.693 \\
46938.828 & 48416.717 &46119.109 &46915.903 &48416.729 \\	
47207.937 & 53832.289 &46915.009 &48416.680 &48416.798 \\	
48416.738 & 53832.321 &46938.790 &53833.462 &53832.321 \\	
53832.311 & 53832.393 &47207.976 &54170.401 &54243.315 \\	
53833.427 & 53833.259 &47218.804&&\\	
54839.502 & 53833.301&53832.251&& \\	
 &53833.336 &53832.321&&\\
 &53833.372 &53832.425&&\\
 &53833.410 &53862.329&&\\
 &54201.345 &54839.535&&\\
 &54243.315 &&&\\
 &54839.453 &&&\\

\hline
\end{tabular}
\end{center}
}
\end{table*}

{\bf NC7}. The O-C diagram for this star suggests that the adopted period is incorrect. However, the amount by which the period is incorrect depends on the cycle counting. In Fig. \ref{OCDIAGS} we show three possibilities. There is no convincing evidence for a secular period change.

{\bf NC11}. Based on the first two points of the O-C diagram Nemec et al. (1995) could not argue in favour of a period change.  The latest group of O-C data however make a rather convincing case for a period decrease at a rate of -0.101 d/Myr. 

{\bf NC13}. Nemec et al. (1995) found that the period of this star is increasing. However we find no evidence of an upward parabola on the O-C diagram but rather a downward one implying a period that
decreases at a rate of -0.011 d/Myr.
Nevertheless the open circles on Fig. \ref{OCDIAGS} show how fragile that conclusion might be. A miscounting of one cycle would imply that the period does not to change at all or changes at a different rate. 

{\bf NC14}. Like NC7, depending upon the cycle counting, three possibilities for the secular behaviour of the period are shown in Fig. \ref{OCDIAGS}. If the parabolas are real then the period change rates are -0.008 or -0.0417 d/Myr for the two presented cases. However, given the uncertainties in the times of maximum light, three straight lines of different slopes are also possible.

{\bf NC15}. Although other possibilities are indicated by the O-C diagram of this star in Fig. \ref{OCDIAGS}, it seems that the parabolic fit is a reasonable
solution. In that case the rate of period decrease would be -0.084 d/Myr.

All the above period change rates suggested by the parabola fits imply new periods
that differ from the ephemerides by about $\times 10^{-7}$ days or less,  
and therefore the new periods cannot be confirmed based exclusively on the data of the present paper.

\subsection{The Blue Stragglers}

\begin{figure} 
\includegraphics[width=8.cm,height=8.cm]{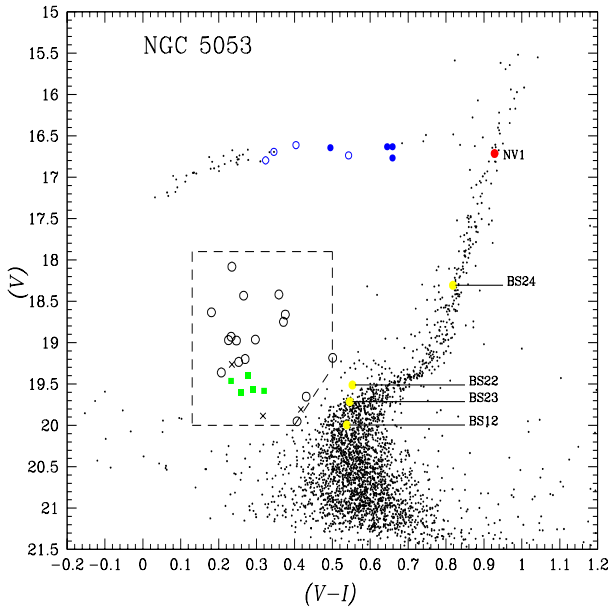}
\caption{Variable stars in the colour-magnitude plane of NGC~5053. Solid blue circles represent RRab stars and open blue circles represent RRc stars. The region enclosed by dashed lines is
the blue straggler region in NGC~6366 according to Harris (1993) but adapted
 to the corresponding brightness and colour of NGC~5053. Open black
circles are previously known BS stars and crosses are newly identified BS stars BS29, BS30 and BS31. Solid green squares
correspond to the five SX Phe stars known to Nemec et al. (1995). The solid yellow circles correspond to stars BS12, BS22, BS23, BS24 previously identified as BS stars which we do not confirm as such. A new RGV, NV1, is shown as a solid red circle.}
    \label{VII}
\end{figure}

There are 28 blue stragglers (BS) reported in NGC~5053, 26 of which
are identified
by Nemec \& Cohen (1989) and Nemec et al. (1995). Five of these BS are the
SX Phe variables discussed in $\S$ \ref{PERSXPHE}. Three BS stars were found by Sarajedini \& Milone (1995) but their star number 25 is the same as BS25, thus
to avoid confusion we shall include Sarajedini \& Milone SM26 and SM27 as
BS27 and BS28 respectively. We have paid special attention to the identification of each of the BS in our images and our data file collection since some of them lie in highly crowded regions. We have correctly found the corresponding $V$, $r$ and $I$ 
files for 25 of the 28 BS since BS20 lies off the field of our images and
BS16 is very near to the image border. BS28 very unfortunately falls on a column
of bad pixels on the CCD chip in the $I$ reference image, and therefore 
we only have $V$ and $r$ light curves. The identified stars are then 
highlighted in Fig. \ref{VII} as solid yellow circles, open 
black circles or solid green squares (see caption). 
The region enclosed by dashed lines in Fig. \ref{VII} is the blue straggler region according
to Harris (1993) for the cluster NGC~6366, but adapted to the brightness and colour of NGC~5053.
21 of the 25 measured BS stars fall within the region confirming their BS nature.
The four BS stars that do not fall in the BS region are BS12, BS22, BS23 and BS24 whose $(V-I)$
colour places them on the main sequence turn off region or on the RGB. They are plotted as
solid yellow circles for easy identification. Through our dedicated effort towards the identification of these stars,
we conclude that these four stars are not likely to be Blue Stragglers.

On the other hand we have found three stars in the BS region not previously identified as BS stars. These are shown as crosses in Fig. \ref{VII} and, to continue with the adopted name convention, in
what follows we shall refer to them as BS29, BS30 and BS31.

\section{Search for new variables}

\begin{figure} 
\includegraphics[width=8.cm,height=8.cm]{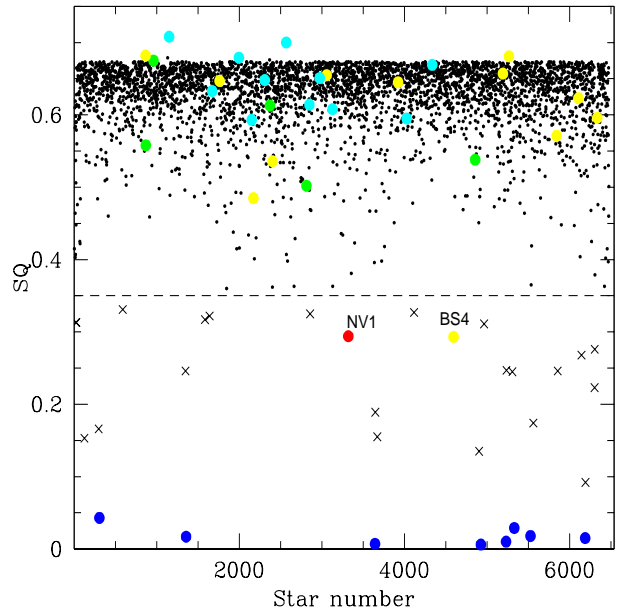}
\caption{SQ parameter distribution for all stars with $V$ light curves in the field of NGC~5053. Light curves for stars below the SQ = 0.35 dashed 
line were individually explored, and only two were classified as 
variables; NV1 (solid red circle) and BS4. Solid blue circles correspond to the known RR Lyrae stars. Solid green circles are the five known SX Phe stars. Solid turquoise circles are non-variable BS stars. Solid yellow circles are the variable BS stars.} 
    \label{SQdiag}
\end{figure}

All the $V$ light curves of the nearly 6500 stars measured in each of the 151 images available were analysed by the phase dispersion minimisation approach (Burke et al. 1970; Dworetsky 1983).
In this analysis the light curve is phased with numerous test periods within a given range. For each period the dispersion parameter SQ is calculated. When SQ is at a minimum, the corresponding period is the best-fit period for that light curve. Bona fide variable stars should have a value of SQ below a certain threshold.
Similar analysis has been used and described in detail in 
previous papers (e.g. Arellano Ferro et al. 2008a;b, 2006). In Fig. \ref{SQdiag}
the distribution of the SQ parameter for the whole sample of stars is shown. As expected the RR Lyrae stars (solid blue circles) have the smallest values of SQ. However it should be noted that the five known
SX Phe stars (solid green circles) all have large values of SQ despite their variability. Likewise, the BS stars  (solid turquoise circles) present large values of SQ, and hence their possible variability cannot be
ruled out by the SQ method. We
will analyse the BS stars in $\S$ \ref{BSsection}.
The $V$ light curves for all stars with
SQ $<$ 0.35 were visually inspected (black crosses) and only two are convincing variables; NV1 (solid red circle) and BS4. Their periodicities will be discussed below.

\begin{figure} 
\includegraphics[width=8.cm,height=8.cm]{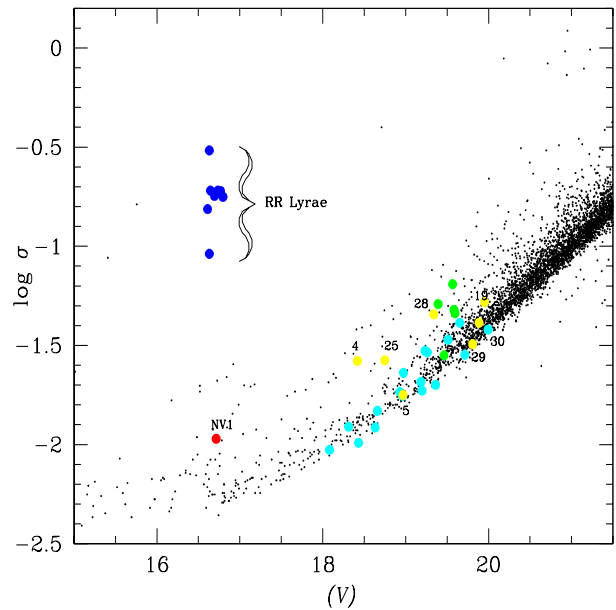}
\caption{Logarithm of the standard deviation of the mean (log $\sigma$) as a function of the mean magnitude $(V)$. Stars above the main
cluster of points are good candidate variables, although small amplitude variability may be found among stars with small log $\sigma$. Solid blue circles are RR Lyrae stars. Solid green circles are the five known SX Phe stars. Solid turquoise circles are non-variable BS stars and 
labelled solid yellow circles are variable BS stars. The new RGB 
variable is shown
as a solid red circle.} 
    \label{ERRdiag}
\end{figure}

Fig. \ref{ERRdiag} shows the logarithm of the standard deviation of the mean (log $\sigma$) as a function of mean magnitude. Stars with large dispersion for a given magnitude are good candidates to be variable. This figure, in combination with Fig. \ref{SQdiag} can be used to identify new variables. In Figs. \ref{SQdiag} 
and \ref{ERRdiag} we also plot the RR Lyrae stars, SX Phe stars, BS stars and the new variables
that will be discussed in detail in the forthcoming sections.

\subsection{Variability among BS stars}
\label{BSsection}

Apart from the 5 stars identified as SX Phe among the BS stars, Nemec et al. (1995) suspected 
variability in the stars BS12, BS21 and BS25.
We explored each BS star light curve in our collection and found clear
indications of variability in some of them. In what follows we will discuss 
their periodicities and in Table \ref{BSdata} we summarise the properties
of the whole collection of confirmed and new BS stars. 
It should be
stressed however, that variables in the BS region often show multiple frequencies and small
amplitudes, e.g. SX Phe stars, and that our light curve sampling is far from ideal
 to establish complicated frequency patterns; although the time span is three years,
the inter-run gaps are large and hence the window function is complicated. This problem was also a limitation of the period determination by Nemec et al. (1995),
hence in depth study of
the frequency content of the variable BS stars, including the known SX Phe stars, long multisite observing campaigns would be required.
Thus we aim to find/confirm the variability among the BS stars and to determine the
main periodicity, although in a few cases, we are able to detect a second frequency. 

The BS stars have mean $V$ magnitudes between 18 and 20. In this range 
the magnitude errors are between 0.01 and 0.04 mag (see Fig. \ref{ERRdiag}). Keeping this in mind, we have 
classified the BS stars as variable, suspect variable and non-variable. 
We confirm the variability of BS25 suspected by Nemec et al.
(1995), and we also find variability in BS4, BS5,
 BS19, BS28, and in the newly discovered BS stars BS29 and BS30. 
In Fig. \ref{ERRdiag}, known and new BS stars are shown as
solid turquoise circles except for the variable ones that are labelled and represented by solid yellow circles.
We describe below our attempts to find their periodicities.

Our approach to determining the periodicities of the BS stars
involves a PDM analysis of the light curve to find a first estimate of the period. Then P4 was run as a confirmation and to search for other frequencies. 
Since the period range in 149 SX Phe stars in Galactic globular clusters is between 0.03 and 0.14d (Rodr\1guez \& L\'opez-Gonz\'alez 2000), we started searching in this range. However, for specific cases shorter or 
longer periods were explored. We address each 
star individually in the following paragraphs.

\begin{figure*} 
\includegraphics[width=17.cm,height=22.cm]{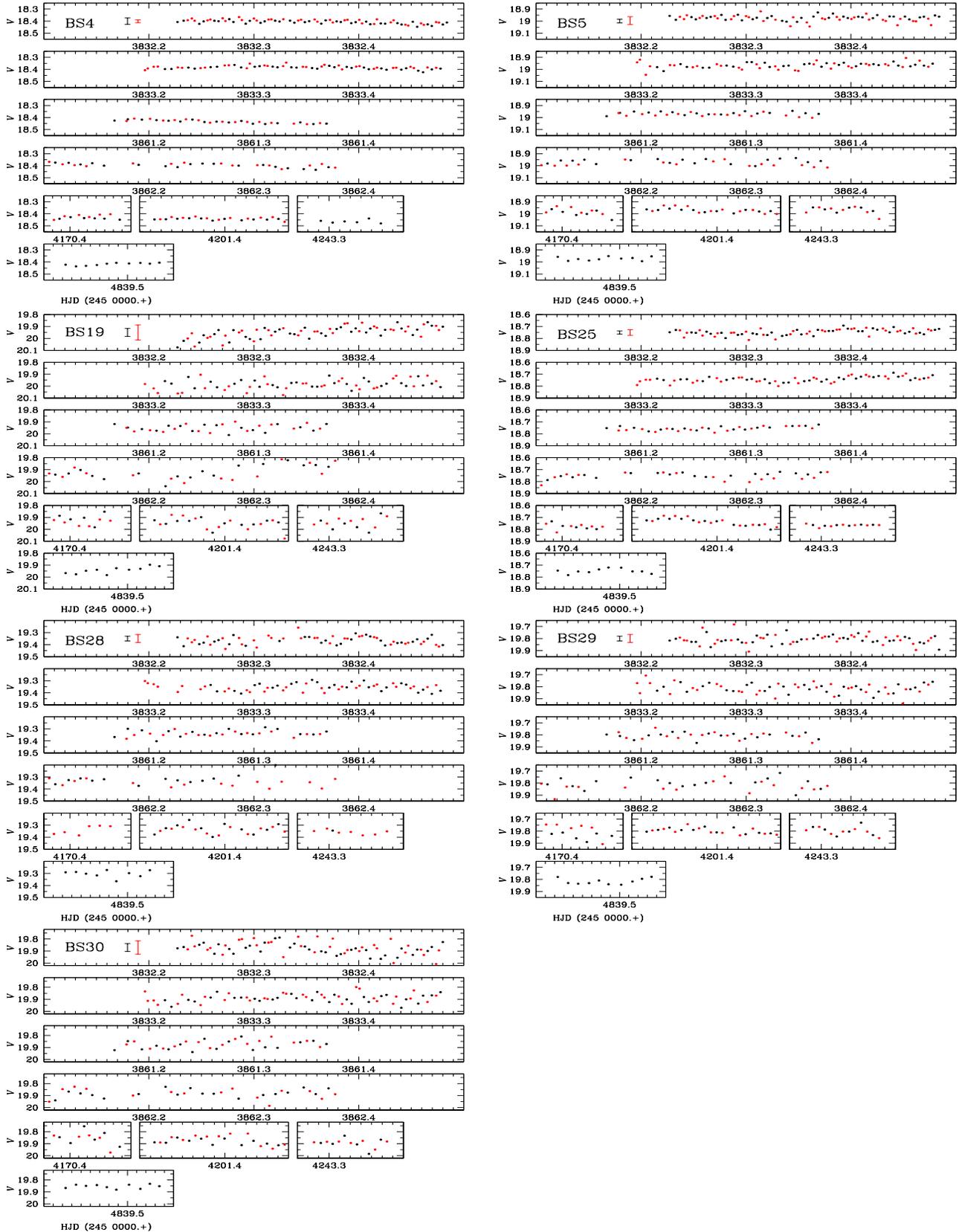}
\caption{$V$ (black circles) and $r$ (red circles) light curves of the variable BS stars in NGC~5053 as obtained in the present work. In order to highlight the variations, the $r$ light curves have 
been offset in magnitude such that the mean $r$ magnitude matches the mean $V$ magnitude for each star. Mean uncertainties for $V$ and $r$ data points are plotted at the start
of the light curve for clarity. The vertical scale is the same for all stars.} 
    \label{VARBS1}
\end{figure*}

\subsubsection{Periodicities in variable BS stars}
\label{BSvari}

The light curves of the variable BS stars are shown in
Fig. \ref{VARBS1}. 

\begin{figure} 
\includegraphics[width=8.cm,height=8.cm]{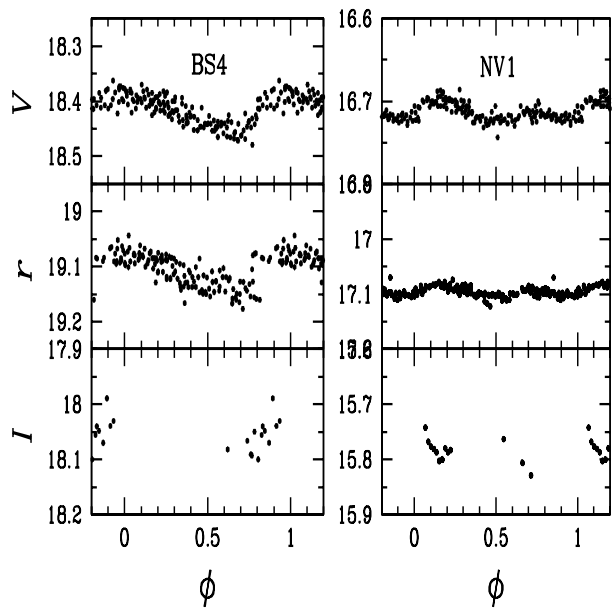}
\caption{Folded $V, r, I$ light curves of two newly identified variables.} 
    \label{2newvars}
\end{figure}

{\bf BS4.} No clear short periods are visible in the light curve and
instead we find long term mean magnitude changes. Two PDM periods are found; 0.07014d and 0.52382d.
While the first period is typical of an SX Phe star, it does not produce
a coherent folded light curve. On the other hand P4 finds a period of 1.10354d which is
nearly twice the longer PDM period. In Fig. \ref{2newvars} the $V, r, I$ light curves are phased with the ephemerides  HJD$_{\rm max}$=245~3833.3057+ 0.52382E, although the period 1.10354d cannot be ruled out.

{\bf BS5.} Variations with a characteristic time of the order of
0.1d are visible in the light curve. In fact the light curve is similar to that of the SX Phe star NC14 in Fig.
\ref{SXstars}. The PDM period is 0.12659d while P4 finds 0.02308, 0.03847 and 0.12678d. One cannot $a priori$ favour any one of these periods,
although we note that if the star is indeed a SX Phe, as suggested by its position on the CMD, then the period 0.03847d
would be the most consistent with the expectation from the PL relationship for SX Phe (see $\S$ \ref{PLsec}).

\begin{table}
\footnotesize{
\begin{center}
\caption[] {\small Blue stragglers in NGC~5053. For known SX Phe we retained the nomenclature NC after Nemec et al. (1995). New BS and not confirmed BS stars are also listed.}
\label{BSdata}
\hspace{0.01cm}
 \begin{tabular}{lccccc}
\hline 
Star&$(V)$ &Variable& $P$&  Type & Notes\\
& (mag.)& &(days) &  & \\
\hline
BS1&18.082 & no &   &    &  \\
BS2&18.432 & no &   &    &  \\
BS3&18.660 & no &   &    &  \\
BS4&18.418 &yes &0.52382& ? & NV \\
BS5&18.969 &yes &0.03847& SX Phe & NV \\
BS6&18.633 & no & &    &  \\
NC7&19.391 & yes &0.03700 &   SX Phe & known \\
BS8&19.233 & no &  &    &  \\
BS9&18.975 & no &  &    &  \\
BS10&18.963& no &  &   &  \\
NC11&19.566  & yes &0.03765  &  SX Phe & known \\
BS12&19.995  & no  &   &  & nBS \\
NC13&19.583  & yes &0.03396  & SX Phe & known \\
NC14&19.463  & yes &0.03411  &  SX Phe & known \\
NC15&19.596  & yes &0.03574  &  SX Phe & known \\
BS16&  &  &  &    & not included \\
BS17&19.361  & no &  &   &  \\
BS18&19.652  & no &  &   &  \\
BS19&19.951  & yes &0.02929  & SX Phe?  & NV\\
BS20&  &  &  &    & not included \\
BS21&18.928  & no &  &    &  \\
BS22&19.512  & yes? &0.07725 & ?  & nBS \\
BS23&19.714  & yes? &0.12102 & ? & nBS\\
BS24&18.307  & no &  &    & nBS \\
BS25&18.748  & yes &0.04521& SX Phe  & NV \\
BS26&19.184  & no &  &    &  \\
BS27&19.808  & no &  &    & SM26 \\
BS28&19.339  & yes &0.04547  & SX Phe   & SM27, NV \\
BS29&19.808  & yes &0.03411 & SX Phe  & NBS, NV \\
BS30&19.885  & yes &0.02535  & SX Phe  & NBS, NV \\
BS31&19.263  & yes? &0.05105  & SX Phe?  & NBS\\

\hline
\end{tabular}
\end{center}
Notes: NV: new variable, nBS: not a Blue Straggler, NBS: new Blue Stragglers, SM26 \& SM27
discovered by Sarejedini \& Milone (1995)
}
\end{table}

{\bf BS19.} The light variation is very evident. The PDM period is 0.07877d. P4 finds a prominent peak at 0.60889d, and after prewhitening, a peak at 0.02929d appears.
This is a similar case to that of BS4 but the phasing with
the longer period 0.60889d does not produce a convincing light curve. 
If the short period is confirmed with more data then the star might be an SX Phe star.

{\bf BS25.} This star is in a crowded region but it can be well measured by our difference imaging technique. In fact, the average magnitude uncertainties for this $V$=18.7 mag star are among the smallest for the BS stars;
0.013 and 0.024 in $V$ and $r$ respectively. Therefore the variations,
although of small amplitude, are likely to be real and we carried out a search for the periodicity. The PDM period is 0.17224 while P4 finds a substantial peak at 0.20821d and a secondary peak at 0.04521. The later period would be 
consistent with the first harmonic of a SX Phe star of the brightness
of BS25, according to the SX Phe $PL$ relation (see $\S$ \ref{PLsec}).

{\bf BS28.} The light curve shows signs of variability. Some epochs are missing because the star falls near a bad pixel column in the chip and then
in some images the photometry is unreliable. The PDM analysis indicates a period of 0.17425d. P4 finds two major peaks at 0.20408d and 0.04547d. The star
is probably an SX Phe.

{\bf BS29.} Clear signs of variability are seen in the light curve. The PDM period is 0.142412d. P4 finds a strong peak at 0.034110d which is nearly one fourth of the PDM period. Prewhitening the main period one finds 0.18477d (see Fig. \ref{POWERS}). It is likely that these two frequencies are real.
The star is a new BS star and likely an SX Phe.

{\bf BS30.} Clear signs of variability are also seen in the light curve of this star. The PDM period is 0.07290d. P4 finds a strong and substantial peak at 0.45045d (A=0.0215 mag). Prewhitening the major frequency one finds 0.02535d. The period 0.45045d does not fold the light curve at all and 
therefore it must be a spurious alias. The shorter period is likely to be correct.  This is a new BS star and very likely an SX Phe.

\begin{figure*} 
\includegraphics[width=17.cm,height=11.cm]{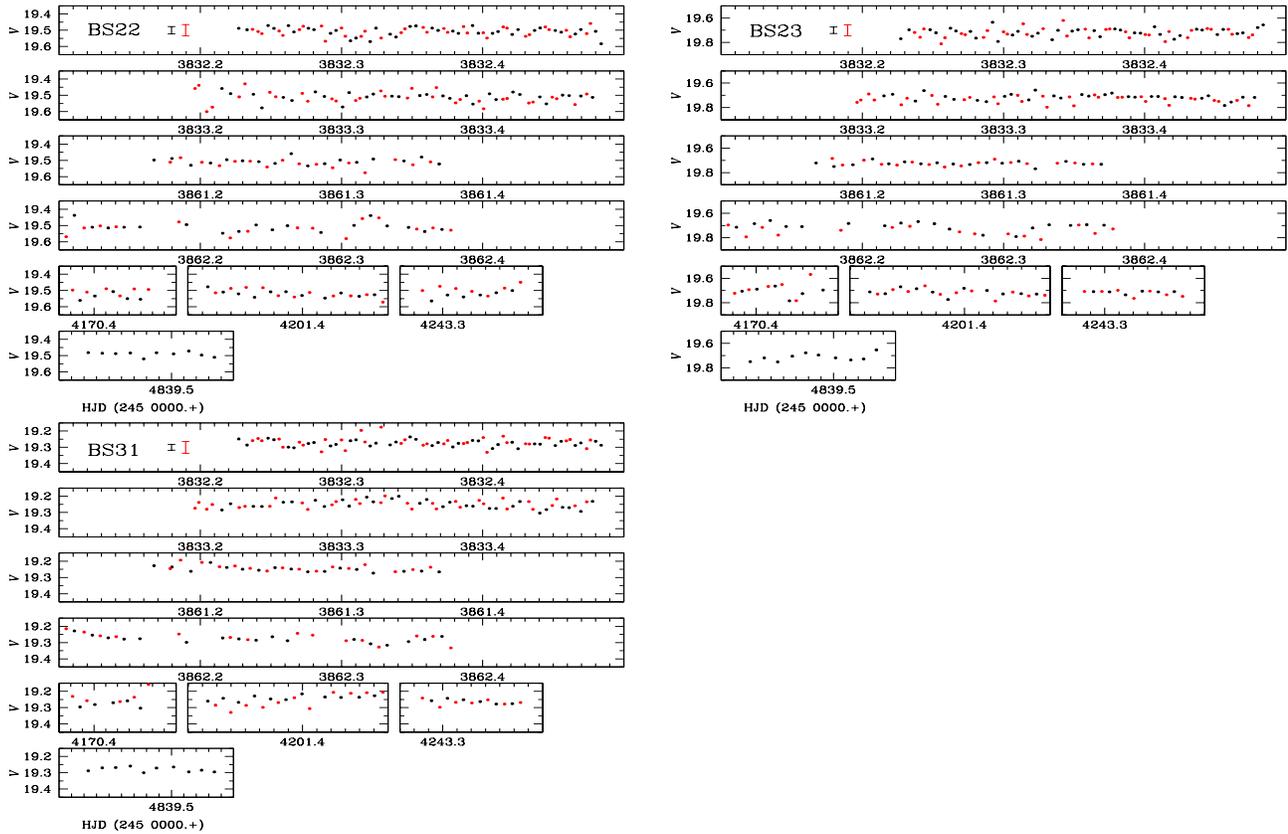}
\caption{$V$ (black circles) and $r$ (red circles) light curve of suspect variable BS stars in NGC~5053 as obtained in the present work. In order to highlight the variations, the $r$ light curves have 
been offset in magnitude such that the mean $r$ magnitude matches the mean $V$ magnitude for each star. Mean uncertainties for $V$ and $r$ data points are plotted at the start
of the light curve for clarity.} 
    \label{SUSVARBS}
\end{figure*}

\subsubsection{Periodicities in suspect variable BS stars}
\label{BSnonvari}

The light curves of the suspect variable BS stars in Fig. \ref{SUSVARBS}.

{\bf BS22.} Considering the average magnitude uncertainties indicated in the figure, mild indications of variability are seen in light curve, especially when the $V$ and $r$ data are combined. We decided to perform a period search in the $V$ data. The PDM period is 0.15580d. P4 finds 0.07725d
and a competing period is 0.16155d. As for the other BS stars,
the removal of the main frequency leaves substantial signal in the power spectrum, suggesting 
the presence of other frequencies which we are unable to determine 
with the available data. The star is not a BS star as it sits on 
the turn off region on its way to the RGB. Its variability will require confirmation with a more appropriate data set.

{\bf BS23.} Similar to BS22, the variations in BS23 are mild. The PDM period is 0.12102d. P4 shows three competing peaks in the main structure of the power spectrum at periods 0.12561, 0.11365 and 0.10287d. The star is not a BS star as it lies in the turn off region (see Fig. \ref{VII}).

{\bf BS31.} The mean magnitude uncertainties for this star are 
0.018 and 0.036 mag in $V$ and $r$ respectively, which implies that the observed variations 
are probably real. The PDM period is 0.02574d, which is too short for an SX Phe star. P4 indicates two competing periods of 0.04556d and 0.05105d. The later is about twice the PDM period and hence the most likely
period. If the variation and period are confirmed then this star might be an SX Phe.

\subsection{The SX Phe $PL$ relation and the distance to NGC~5053}
\label{PLsec}

It is known that SX Phe stars follow a tight relation between their fundamental mode, first overtone periods and their luminosity (McNamara 1995). Due to the mixture of modes, the Period-Luminosity ($PL$) relation
is difficult to determine. Of particular relevance to the present work is the calibration that Jeon et al. (2004) calculated
for seven SX Phe in NGC~5466; $M_V = -3.25 (\pm 0.46) ~log P - 1.30 (\pm 0.06)$ ($\sigma=0.04$), because the distance and iron content of
NGC~5466 ($16.06 \pm 0.09$~kpc and $-1.91 \pm 0.19$; Arellano Ferro et al.
2008a) are almost identical to our values  determined for NGC~5053
($16.75 \pm 0.39$~kpc and $-1.97 \pm 0.16$). A $PL$ dependence on [Fe/H] has been proposed by Nemec et al. (1994) of the form:
$M_V = -2.56 (\pm 0.54) ~log P + 0.36 + 0.32{\rm [Fe/H]}$.

\begin{figure} 
\includegraphics[width=8.cm,height=8.cm]{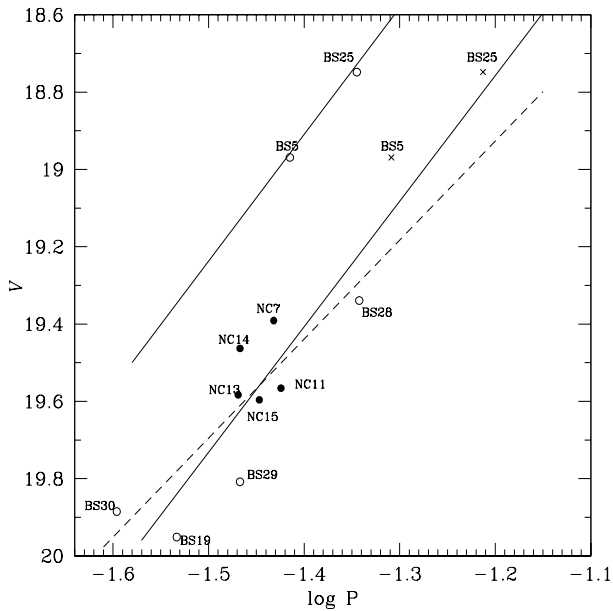}
\caption{Period-Luminosity relation for the SX Phe stars in NGC~5053.
Filled circles represent the known SX Phe stars. Open circles represent the newly identified SX Phe stars among the BS stars. Two positions are shown for BS5 and BS25 for the first overtone period (open circles) and the fundamental period (crosses). The two continuous lines correspond to the first overtone and fundamental mode relations of Jeon et al. (2004)  and the dashed line represents the relation of Nemec et al. (1994). }
    \label{SXMv_P}
\end{figure}

In Fig. \ref{SXMv_P} we have plotted the logarithm of the period (column 4 Table  \ref{BSdata}) vs. the mean magnitude $(V)$ (column 2 Table  \ref{BSdata}) 
for the 5 known SX Phe stars and the BS stars identified as SX Phe stars in $\S$ \ref{BSsection}. The calibrations of Jeon et al. (2004) and Nemec et al. (1994)
are shown as continuous and dashed lines respectively and have been scaled to the case of NGC~5053 using
a distance modulus of 16.125 and $E(B-V)=0.018$. It can be seen that the SX Phe stars in NGC~5053 follow the above two relations.
The most discordant stars would seem to be BS5 and BS25. However, their periods found in $\S$ \ref{BSsection} seem to correspond to the first
 overtone, as suggested by the first overtone calibration of 
Jeon et al. (1994). Their periods 
can be converted to the fundamental mode via the ratio $P_{1H}/P_F = 0.783$ (Jeon et al. 2004), and then the stars also seem to follow the fundamental mode relation (crosses in Fig. \ref{SXMv_P}).

The distribution of the BS stars along the $PL$ relations for SX Phe stars
confirms their SX Phe nature. Since the period for the SX Phe stars in NGC~5466 is better determined than for the SX Phe stars in NGC~5053, rather than attempting an independent calibration of the $PL$ relation, the SX Phe (known and newly discovered) can be used to estimate the distance to NGC~5053.

The application of the $PL$ relations of Jeon et al. (2004) and
of Nemec et al. (1994) lead to the average distances of
$16.36 \pm 0.85$ and $16.08 \pm 0.98$ kpc respectively. Although these determinations of the distance cannot compete in precision with the determination from the RR Lyrae stars, it is rewarding to find that the agreement is good to within the uncertainties.

\subsection{Other new variables}

In Fig. \ref{ERRdiag} it can be observed that above the main stream there are numerous
points which may be variables. An exploration of all of the corresponding data files up to V$\sim 20$ shows that often the large
sigmas are produced by a few occasional spurious measurements, particularly  
in stars near to the border of our images, or near to saturated or to
authentic variable stars, and are not due to intrinsic variability. However, we find one star 
suspected of real variations as addressed below.

{\bf NV1.} On the CDM this star is on the RGB. Its PDM period
of 0.66578d phases the variations well in all filters (Fig. \ref{2newvars}). No evidence of more frequencies was found with the data on hand. P4 suggests two peaks 
at 1.69426d and 0.63921d, none of which phase the light curve as well as the PDM period.

\begin{figure*} 
\includegraphics[width=16.cm,height=12.cm]{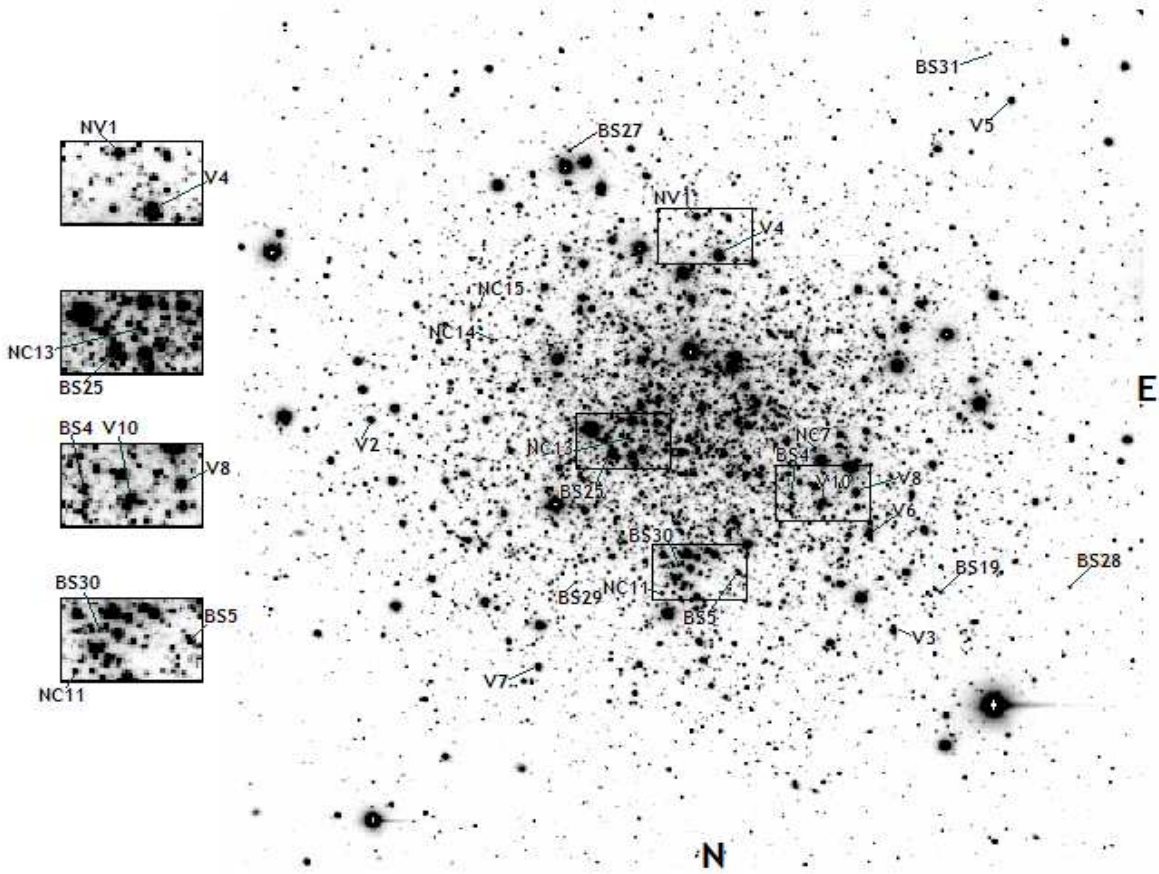}
\caption{V image of NGC~5053 obtained at the Indian Astrophysical Observatory at Hanle, India. The size is approximately 11$\times$11 arcmin$^2$.
All RR Lyrae stars are identified with prefix V, Blue Stragglers with prefix BS, and the five previously known SX Phe stars with prefix NC. The two BS stars 
found by Sarajedini \& Milone (1995) are BS27 and BS28. The three new BS stars found in this work are BS29, BS30 and BS31.
The rest of the Blue Stragglers are clearly identified by Nemec \& Cohen (1989).} 
    \label{ELCUMULO}
\end{figure*}

In Fig. \ref{ELCUMULO} we display a finding chart for the eight RR 
Lyrae stars studied, the five previously known SX Phe stars, 
the variable BS stars, the three new BS stars and the new variable NV1.

\section{Discussion}

\subsection{On the distance to NGC~5053}

To determine the distance to the cluster we have calculated the distance modulus $(A_0 - M_V)$ 
 for each RR Lyrae star, using the absolute magnitudes $M_V(K)$ given in {Tables  ~\ref{fisicosC} and {\ref{fisicosAB}, and the mean magnitudes $A_0$ from {Table ~\ref{foufit}}. We used the extinction ratio 
R=3.2 and the value of $E(B-V)= 0.018$ was adopted from Nemec (2004).
 
We find mean true distance moduli of $16.12 \pm 0.05$ and $16.12 \pm 0.03$, which correspond to distances of $16.71 \pm 0.39$ kpc and
$16.75 \pm 0.26$ kpc for the RRc and RRab stars respectively. 
The uncertainties quoted for these values correspond to the standard deviation of the mean. 
These values of the distance are consistent with the mean luminosity of the RR Lyrae stars in the LMC $V_o=19.064 \pm 0.064$ (Clementini et al. 2003) and a distance modulus of $18.5 \pm 0.1$ mag for the LMC (Freedman et al. 2001; van den Marel et al. 2002; Clementini et al. 2003). 

In his catalogue of 
parameters of globular clusters, Harris (1996) lists a distance of 16.4 kpc for NGC~5053, which comes from the assumption of  $V$(HB)=16.65 (Sarajedini \& Milone 1995), [Fe/H]=$-2.29$, R=3.0, $E(B-V)=0.04$ and the relation $M_V=0.15{\rm [Fe/H]} + 0.80$. His calculation is mostly sensitive to the values of [Fe/H] and $E(B-V)$. If the calculation is repeated with the value [Fe/H]=$-1.97$, and depending upon 
the assumed value of $E(B-V)$ between 0.018 and 0.04 mag, one obtains a distance in the range 16.04 to 16.53 kpc.
No significant differences are found if a reasonably different $M_V$-[Fe/H] relation is used.

The Fourier results for the distance are independent of the iron abundances however, they seem to support the estimate made using low metallicity.

\subsection{On the age of NGC~5053}

We have estimated the age of the cluster by a direct comparison of the cluster's CMD with the theoretical isochrones of VandenBerg, Bergbusch \& Dowler (2006)
as shown in Fig. \ref{AGE}. To properly align the 
theoretical isochrones with the observed distribution of stars, we aligned the most 
blue  point on the isochrones with the Turn Off (TO) point. Indeed, the main 
source of uncertainty in this procedure is in the location of the TO point. In fact, Rosenberg et al. (1999) report that the TO point of NGC~5053 is at $V=20.00 \pm 0.06$ and 
$(V-I) = 0.545 \pm 0.004$. This TO point is shown in Fig. \ref{AGE} as a solid white
circle and indicates that the best fit model is for an age of $\sim$ 16 Gyrs, as
shown in the bottom panel.
The age seems much too large  when compared to the most recent differential calculations of globular cluster ages (as we shall see below) which further implies
that the adopted TO point is too red.

We have attempted our own estimate of the location of the TO point. We adopted $V=20.0$, but for the 
$(V-I)$ value we proceeded as follows; the mean uncertainties in our photometry at $V =20.0$
and $I =19.5$ are 0.04 and 0.05 mag respectively, and hence the uncertainty in $(V-I)$ is
$\sim$ 0.06 mag, which is represented by the horizontal error bars shown in Fig. \ref{AGE}. To allow for this uncertainty in the colour, we have shifted the TO point
from the bluest point of the bulk of stars at $V =20.0$,
by 0.06 mag redwards. In the top panel of Fig. \ref{AGE} we have fixed 
the isochrones on 
this new TO point as we did in the bottom panel. The best fit of the RGB would be 
for a model of age between 12 and 14 Gyrs. The best fits for 12 Gyrs and 14 Gyrs 
would be obtained for shifts of the TO from the bluest point of 0.05 and 0.09 mag respectively. A linear interpolation in between these two models for 0.06 mag gives an age of 12.5 Gyrs. The uncertainty attached to this estimate is related to the uncertainty
in the location of the TO point due to uncertainties in the photometry. Shifting the TO point $\pm 0.03$ mag would correspond to an uncertainty of about $\pm 2.0$ Gyrs in the age estimation. 

A differential version of the vertical and horizontal methods for globular cluster age determination for a family of Galactic globular clusters 
can be found in the work of Rosenberg et al. (1999). NGC~5053 is included in 
the cluster sample used by these authors who estimate an average age for the cluster of $-0.9 \pm 0.07$ Gyrs relative to the absolute age of a group of coeval clusters
assumed to have an age of 13.2 Gyrs as in Carretta et al. (2000). In other words,
their estimate for 
the age of NGC~5053 is $12.3 \pm  0.7$ Gyrs. These authors consider NGC~5053 to be
a member of the coeval group. 

The CMD in Fig. \ref{AGE} is comparable to the CMD of the cluster published by Rosenberg et al. (2000), which is to be expected since we have used the same set of standard stars. The Rosenberg et al. (2000) diagram was used by Rosenberg et al. (1999) to estimate the age of the cluster.
The horizontal fiducial lines in Fig. \ref{AGE} from the bottom upwards correspond to the TO point,
the fiducial point on the RGB 2.5 mag above the TO point (RGB$_{2.5}$), and the level of the HB respectively.
We have estimated the age sensitive vertical 
and horizontal parameters  $\Delta V^{HB}_{\rm TO} = {\rm TO} - {\rm HB} = 3.35 \pm 0.05$ mag
and $\delta (V-I)_{2.5} = \rm RGB_{2.5} - {\rm TO} = 0.359 \pm 0.050$ mag, that can be compared with those estimated by Rosenberg et al. (1999) as $3.30 \pm 0.08$
and $0.310 \pm 0.007$ respectively. It is clear that vertically our results agree well and hence similar ages would be found. However, due to the 
substantially different horizontal position of the TO point, a large difference is exhibited between the values of the
$\delta (V-I)_{2.5}$ parameter. Despite this discrepancy, our estimate agrees well with their horizontal estimate of the age of $12.6\pm0.5$ Gyrs. This is probably due to the fact that, despite Rosenberg et al.'s (1999) absolute estimation of the 
mean age for the group of
coeval clusters from the horizontal parameter and a comparison with the models of Straniero et al. (1997) and VandenBerg et al. (1990) of 13.1 and 16.4 Gyrs respectively, they adopt an age for the coeval group of 13.2 Gyrs, and then  estimate individual cluster ages differentially.
Our estimate for the age of NGC~5053 agrees well within the uncertainties
with the final average age given by Rosenberg et al. (1999).

Another age estimate for NGC~5053 obtained by employing a differential approach is that of Salaris \& Weiss (2002) who found an age of $10.8\pm0.9$ Gyrs. The most recent differential age calculation for
64 galactic globular clusters was made by Mar\'in-Franch et al. (2009) using four sets of 
stellar evolution libraries. These authors find an average relative age $A_{NGC~5053}/A_{bulk} = 0.96 \pm 0.04$, and using the models of Dotter et al. (2007), they calculated a mean absolute age for the low metallicity clusters
of $12.8\pm0.17$ Gyrs, from which one can calculate an age of $12.29 \pm 0.17$ Gyrs for NGC~5053.

There is reasonable agreement between our result of $12.5 \pm 2.0$ Gyrs with the three independent and more accurate differential age estimates of $12.3 \pm  0.7$ Gyrs (Rosenberg et al.  1999),
 $10.8\pm0.9$ Gyrs (Salaris \& Weiss 2002) and $12.29 \pm 0.17$ Gyrs (Mar\'in-Franch et al. 2009).

\begin{figure} 
\includegraphics[width=8.cm,height=12.cm]{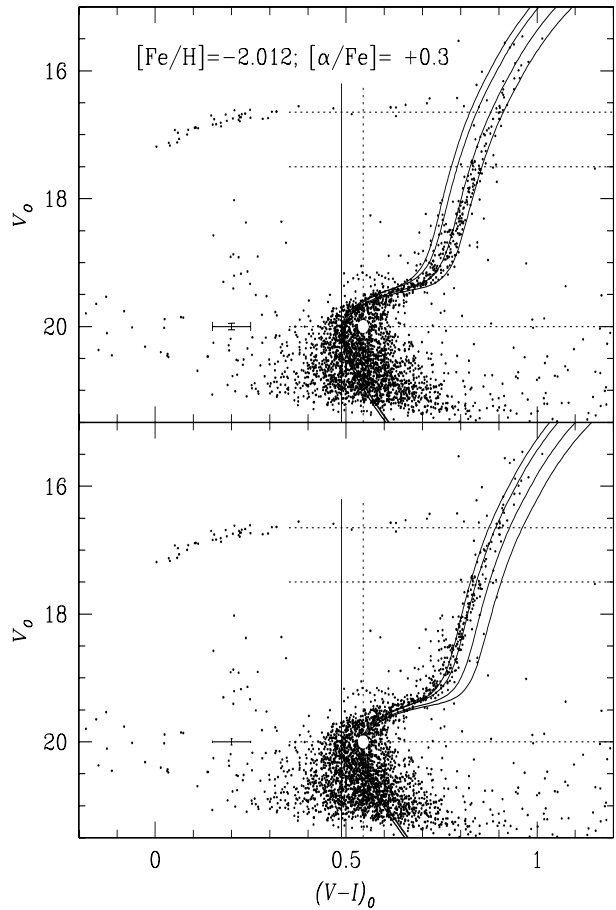}
\caption{Isochrones from models with [Fe/H]= $-2.012$ and [$\alpha$/Fe]=0.3 of 
VandenBerg, Bergbusch \& Dowler (2006). The ages of the isochrones are identified at the RGB from
right to left as 12, 14, 16 and 18 Gyrs. The bottom panel shows the models aligned
with the turn off point (TO) of Rosenberg et al. (2000) (solid white circle). In the top panel the models are aligned with the TO point calculated in the present paper. The error bars correspond to the mean uncertainties of the $V$ and $I$ photometry 
at $V=20.0$ mag and $I =19.5$ mag. See text for discussion.}
    \label{AGE}
\end{figure}

\section{Conclusions}

The technique of difference imaging has proven to be a powerful tool in  
providing accurate photometry down to about $V=20$ mag and in finding new variables. In this paper we report the discovery of five SX Phe stars among the BS stars, two of
which are new BS stars, and one new red semiregular variable.

Physical parameters of astrophysical relevance such as: $log~(L/L_{\odot})$, 
 $log~T_{\rm eff}$, $M_V$, $R/R_{\odot}$, $M/M_{\odot}$ and [Fe/H], have been derived for the RR Lyrae stars in NGC~5053
using the Fourier decomposition of their $V$ light curves. Special attention has been paid to the calibration of these parameters onto widely accepted scales.
It has been found that $T_{\rm eff}$ for the RRc stars from the calibration of Simon \& Clement (1993) cannot be reconciled with the theoretical predictions for the 
boundaries of the instability strip. However, the Fourier temperatures for RRab stars from the calibration
of Jurcsik (1998) agree well with the colour temperatures, and they imply
radii and masses comparable to those derived from pulsational models.

The average iron abundance [Fe/H] = $-1.97 \pm 0.16$ is found from the RRab and RRc stars. This value may appear too large for a cluster often considered the most metal deficient in the Galactic halo. However, this value is in good agreement,
 within the uncertainties, with the most recent accurate estimates from spectroscopic data.

A distance of $16.7 \pm 0.3$ kpc is found from the Fourier approach to the determination of $M_V$ for the RR Lyrae stars. This value is in agreement, within
uncertainties, with predictions from the $M_V$-[Fe/H] relation for 
values of [Fe/H] $\sim -2.0$.

We report five new SX Phe stars and show that they follow the period luminosity relationship of SX Phe stars in NGC~5466, a cluster very similar in age and metallicity to NGC~5053. If the $PL$ calibration
is applied to the SX Phe in NGC~5053, then the estimated distance is found 
in good agreement with the distance from the RR Lyrae stars.
  
Our $(V-I)$ CMD and the isochrones of VandenBerg, Bergbusch \& Dowler (2006)
indicate an age of $12.5 \pm 2.0$ Gyrs for the cluster, which is in good agreement with independent differential age estimations.

\section*{Acknowledgments}

We are grateful to the support astronomers of IAO, at Hanle and CREST (Hosakote), for their efficient help while acquiring the data and to the referee for the detailed revision and very 
useful suggestions. The help of Victoria Rojas with the production of Fig. \ref{ELCUMULO} is
acknowledged with gratitude. We acknowledge support from the DST-CONACYT collaboration project and the DGAPA-UNAM grant through project IN114309 at several stages of the work. AAF is grateful to the IIA for their warm hospitality. DMB is
thankful to the Instituto de Astronom\1a of the Universidad Nacional Aut\'onoma de M\'exico for their hospitality. This work has made a large use of the SIMBAD and ADS services, for which we are thankful.


\begin{thebibliography}{99}
\bibitem{} Alard C., 2000, A\&AS, 144, 363
\bibitem{} Alard C., Lupton R.H., 1998, ApJ, 503, 325
\bibitem{} Arellano Ferro A., Ar\'evalo M. J., L\'azaro C., Rey M., Bramich D. M., Giridhar S., 2004, RevMexAA, 40, 209
\bibitem{} Arellano Ferro, A., Rojas L\'opez, V, Giridhar, S., Bramich, D.M., 2008a, MNRAS, 384, 1444
\bibitem{} Arellano Ferro, A., Giridhar, S., Rojas L\'opez, V, Figuera, R, Bramich, D.M., Rosenzweig, P., 2008b, RevMexAA, 44, 365
\bibitem{} Armandroff, T.E., Da Costa, G.S., Zinn, R.J., 1992, AJ, 104, 164 
\bibitem{} Barnes, T. G., Evans, D. S.,  Moffett, T. J., 1978, MNRAS, 183, 285
\bibitem{} Bell, R.A., Gustafsson, B., 1983, MNRAS, 204, 249
\bibitem{} Benedict G. F., McArthur, B. E., Fredrick, L. W., et al. 2002, ApJ, 123, 473
\bibitem{} Bono G., Caputo F., Castellani V., Marconi M., 1995, AJ, 110, 2365
\bibitem{} Bono, G., Caputo, F., Castellani, V., Marconi, M., Storm, J., Degl’Innocenti, S., 2003, MNRAS, 344, 1097 
\bibitem{} Bramich D. M., Horne K., Bond I. A., Street R. A., Cameron A. C., Hood B., Cooke J., James D., Lister, T. A., Mitchell D., Pearson K., Penny A., Quirrenbach A., Safizadeh N., Tsapras Y., 2005, MNRAS, 359, 1096
\bibitem{} Bramich, D. M. 2008, MNRAS, 386, L77
\bibitem{} Burke,E.W., Rolland, W.W., Boy, W.R., 1970, JRASC, 64, 353
\bibitem{} Cacciari, C., Clementini, G., 2003, LNP, 635, 105
\bibitem{} Cacciari, C., Corwin, T.M., Carney, B.W., 2005, AJ, 129, 267 
\bibitem{} Carretta, E., Gratton, R. G., Clementini, G., Fusi Pecci, F., 2000, ApJ, 533, 215
\bibitem{} Castelli, F., 1999, A\&A, 346, 564
\bibitem{} Chaboyer, B., 1999, in Post-Hipparcos Cosmic Candles, eds. A. Heck \& F. Caputo (Dordrech: Kluwer), p. 111
\bibitem{} Clement C. M., Muzzin, A., Dufton, Q., Ponnampalam, T., Wang, J., Burford, J., Richardson, A., Rosebery, T., Rowe, J., Hogg, Sawyer-Hogg H., 2001, AJ, 122, 2587
\bibitem{} Clementini, G.,  Gratton R. G., Bragaglia, A., et al. 2003, AJ, 125, 1309
\bibitem{} Dotter, A., Chaboyer, B., Jevremovi\'c , D., Baron, E., Ferguson, J. W., Sarajedini, A., Anderson, J. 2007, AJ, 134, 376
\bibitem{} Cox, A. N., Hudson, S. W., Clancy, S. P., 1983, ApJ, 266, 94
\bibitem{} Dworetsky, M. M., 1983, MNRAS, 203, 917
\bibitem{} Di Fabrizio, L., Clementini, G., Maio, M., Bragaglia, A., Carretta, E., Gratton R. G., Montegriffo, P., Zoccali, M., 2005, A\&A, 430, 603
\bibitem{} Freedman, W.L., Madore, B.F., Gibson, B.K. et al. 2001, ApJ, 553, 47
\bibitem{} Gratton R. G., Bragaglia, A., Clementini, G., Carretta, E., Di Fabritzio, L., Maio, M., Taribello, E., 2004, A\&A, 221, 937
\bibitem{} Geisler, D., Piatti, A.E., Clari\'a, J.J., and Minniti, D. 1995,
          AJ 109, 605
\bibitem{} Harris, H.C., 1993, AJ, 106, 604
\bibitem{} Harris, W.E., 1996, AJ, 112, 1487
\bibitem{} Jeon, Y.-B., Lee M.G., Kim S.-L, Lee H., 2004, AJ, 128, 287.
\bibitem{} Jurcsik, J., Acta Astron., 1995, 45, 6653
\bibitem{} Jurcsik, J., 1998, A\&A, 333, 571
\bibitem{} Jurcsik, J., Kov\'acs G., 1996, A\&A, 312, 111
\bibitem{} Kinman, T.D., 2002, Inf. Bull. Var. Stars, No. 535
\bibitem{} Kov\'acs, G., 1998, Mem. Soc. Astron. Ital., 69, 49.
\bibitem{} Kov\'acs, G., 2002, in ASP Conf. Ser. 265, $\omega$ Centauri: A Unique Window into Astrophysics. Eds. van Leeuwen, F., Hughes, J., Pioto, G., (San Francisco; ASP), p. 163
\bibitem{} Kov\'acs, G., Kanbur, S.M., 1998, MNRAS, 295, 834 
\bibitem{} Kov\'acs, G., Walker, A.R., 2001, A\&A, 371, 579
\bibitem{} L\'azaro C., Arellano Ferro A., Ar\'evalo M. J., Bramich D. M., Giridhar S., Poretti E., 2006, MNRAS, 372, 69
\bibitem{} Lenz, P., Breger, M. 2005, {\it Communications in Asteroseismology}, 146, 43 
\bibitem{} McNamara, D.H., 1995, AJ, 109, 1751.
\bibitem{} Marconi, M., Nordgren, T., Bono, G., Schnider, G., Caputo, F., 2005, ApJ, 623, 133 
\bibitem{} Mar\1n-Franch, A., Aparicio, A., Piotto, G., Rosenberg, A., Chaboyer, B., Sarajedini, A., Siegel, M., Anderson, J., Bedin, L. R., Dotter, A., Hempel, M., King, I., Majewski, S., Milone, A. P., Paust, N., Reid, I. N., 2009, ApJ, 694, 1498
\bibitem{} Mannino, G., 1963, Pub. Obs. Bologna, 8, 12
\bibitem{} Montegriffo, P., Ferraro, F. R., Origlia, L., Fusi Pecci, F., 1998, MNRAS, 297, 872
\bibitem{} Morgan, S.M., Wahl, J.N., Wieckhorst, R.M., 2007, MNRAS, 374, 1421
\bibitem{} Nemec, J.M., 1989, in {\it The Use of Pulsating Stars in Fundamental Problems of Astronomy}, IAU Coll. 111, ed. E.G. Schmidt, p. 215
\bibitem{} Nemec, J.M., 2004, AJ, 127 2185.
\bibitem{} Nemec, J.M., Cohen, J.G., 1989, ApJ, 336, 780
\bibitem{} Nemec, J.M., Mateo, M., Schombert, J.M., 1995, AJ, 109, 618
\bibitem{} Nemec, J.M., Linnell Nemec, A.F., Lutz, T.E., 1994, AJ, 108, 222.
\bibitem{} Nemec, J.M., Mateo, M., Burke, M., Olszewski, E. W., 1995, AJ, 110, 1186
\bibitem{} Rodr\1guez, E., L\'opez-Gonz\'alez, M. J., 2000, A\&A, 359, 597
\bibitem{} Rosenberg, A., Saviane, I., Piotto, G., Aparicio, A., 1999, ApJ, 118, 2306
\bibitem{} Rosenberg, A., Aparicio, A., Saviane, I., Piotto, G. 2000, A\&AS, 145, 451
\bibitem{} Rutledge, G.A., Hesser, J.E., Stetson, P.B., 1997, PASP, 109, 907 
\bibitem{} Salaris M., Chieffi A., Straniero O., 1993, ApJ, 414, 580 
\bibitem{} Salaris M., Weiss, A., 2002, A\&A, 388, 492
\bibitem{} Sarajedini, A., Milone, A.A.E., 1995, AJ, 109, 269
\bibitem{} Sekiguch, M., Fukugita, M., 2000, AJ, 120, 1072
\bibitem{} Simon N. R., Clement C. M., 1993, ApJ, 410, 526
\bibitem{} Stetson, P. 2000, PASP, 112, 773
\bibitem{} Straniero, O., Chieffi, A., Limongi, M. 1997, ApJ, 490, 425
\bibitem{} Suntzeff, N.B., Kraft, R.P., Kinman, T.D. 1988, AJ 95, 91
\bibitem{} van Albada, T.S., Baker, N., 1971, ApJ, 169, 311
\bibitem{} VandenBerg, D.A., Bergbusch, P.A., Dowler, P.D., 2006, ApJS, 162, 375
\bibitem{} VandenBerg, D.A., Bolte, M., Stetson, P.B., 1990, AJ, 100, 445
\bibitem{} VandenBerg, D.A., Clem, J.L., 2003, AJ, 126, 778
\bibitem{} van den Marel, R.P., Alves, D.R., Hardy, E., Suntzeff, N.B., 2002, AJ, 124, 2639
\bibitem{} Zinn R., 1985, ApJ, 293, 424
\bibitem{} Zinn R., West, M.J., 1984, ApJS, 55, 45

\end{thebibliography}
\end{document}